\documentclass[onecolumn,amsmath,amssymb,pra,superscriptaddress,a4paper,floatfix,showpacs,showkeys]{revtex4}
\usepackage{graphicx}
\usepackage{mathptmx}      
\begin{document}

\title{Event-based simulation of neutron interferometry experiments\footnote{Accepted for publication in QUANTUM MATTER}}

\author{H. De Raedt}
\email{h.a.de.raedt@rug.nl}
\affiliation{%
Department of Applied Physics,
Zernike Institute for Advanced Materials,
University of Groningen, Nijenborgh 4, NL-9747 AG Groningen, The Netherlands
}%
\author{F. Jin}
\email{f.jin@fz-juelich.de}           
\affiliation{%
Institute for Advanced Simulation, J\"ulich Supercomputing Centre \\
Forschungszentrum J\"ulich, D-52425 J\"ulich, Germany
}%
\author{K. Michielsen}
\email{k.michielsen@fz-juelich.de}           
\affiliation{%
Institute for Advanced Simulation, J\"ulich Supercomputing Centre \\
Forschungszentrum J\"ulich, D-52425 J\"ulich, Germany
}%
\affiliation{%
RWTH Aachen University, D-52056 Aachen,
Germany
}%
\date{\today}

\begin{abstract}
A discrete-event approach, which has already been shown to give a
cause-and-effect explanation of many quantum optics experiments,
is applied to single-neutron interferometry experiments.
The simulation algorithm yields a logically consistent description in terms of
individual neutrons and does not require the knowledge of the solution of a wave equation.
It is shown that the simulation method reproduces the results of several single-neutron interferometry experiments,
including experiments which, in quantum theoretical language, involve entanglement.
Our results demonstrate that classical (non-Hamiltonian) systems can exhibit correlations
which in quantum theory are associated with interference and entanglement,
also when all particles emitted by the source are accounted for.
\end{abstract}


\maketitle
\tableofcontents

\newcommand\sumprime{\mathop{{\sum}'}}
\newcommand\QT{quantum theory}
\newcommand\T{t}
\newcommand\R{r}
\newcommand\RN{{\cal R} }

\section{Introduction}\label{Introduction}

Quantum theory has proven extraordinarily powerful for describing the
statistical properties of a vast number of laboratory experiments.
Conceptually, it is straightforward to use the quantum theoretical formalism to
calculate numbers that can be compared with experimental data, at least if
these numbers refer to statistical averages.
However, a fundamental problem appears if an experiment provides
access to the individual events that contribute to the statistical average.
Prime examples are the single-electron two-slit experiment~\cite{TONO98},
neutron interferometry experiments~\cite{RAUC00} and
similar experiments in optics where the click
of the detector is identified with the arrival of a single photon~\cite{GARR09}.

Although quantum theory provides a recipe to compute the frequencies for observing events
it does not account for the observation of the individual detection events themselves~\cite{HOME97,BALL03}.
From the viewpoint of quantum theory, the central issue is how it can be that
experiments yield definite answers.
As stated by Leggett~\cite{LEGG87}: ``In the final analysis,
physics cannot forever refuse to give an account of how it is
that we obtain definite results whenever we do a particular measurement''.
For a recent review of various approaches to the quantum measurement problem and an explanation of it within
the statistical interpretation, see Ref.~\cite{NIEU11b}.

Perhaps the most simple and clear demonstration
of the fundamental nature of this problem
is provided by two-path interference experiments with electrons, photons, or neutrons.
According to Feynman, the observation that the interference patterns are built up event-by-event
is ``impossible, absolutely impossible to explain in any classical way
and has in it the heart of quantum mechanics. In reality it is the {\sl only} mystery.''~\cite{FEYN65}.

Reading ``any classical way'' as ``any classical Hamiltonian mechanics way'',
Feynman's statement may be difficult to dispute.
However, taking a broader view by allowing for dynamical systems that are outside the realm of classical
Hamiltonian dynamics, it may be possible to model the gradual appearance of interference patterns
through a discrete-event simulation that does not make reference to wave theory.
This is precisely the approach taken in the present paper which
is not about interpretations or extensions of quantum theory (see Ref.~\cite{RAUC00} for an overview)
but adopts a new paradigm~\cite{RAED05b,RAED05c,MICH11a} to
deal with the fact that experiments yield definite results.

\begin{figure*}[t]
\begin{center}
\includegraphics[width=14cm ]{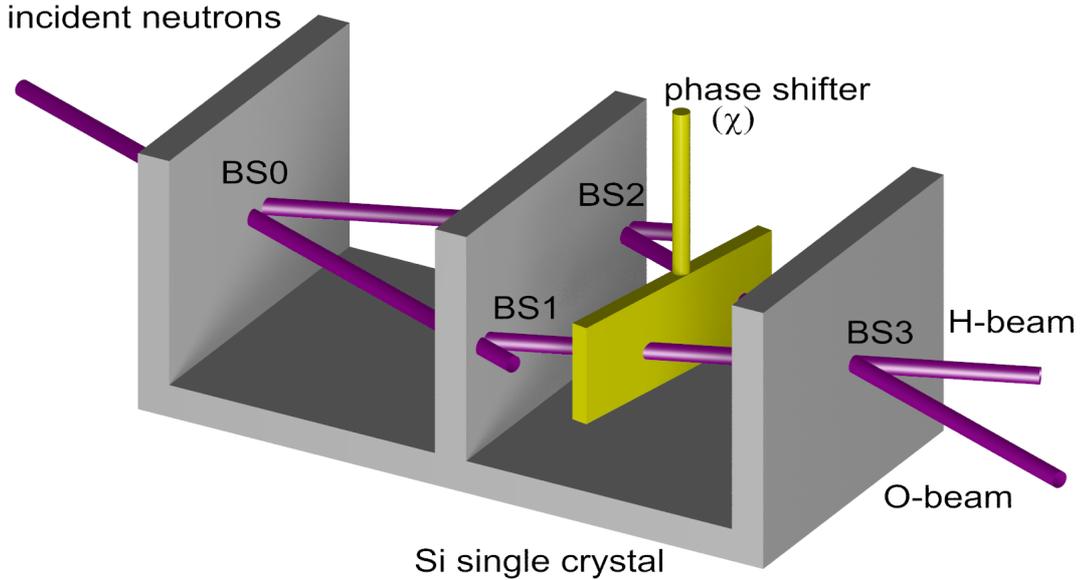}
\caption{%
Picture of the perfect crystal neutron interferometer~\cite{RAUC74a}.
BS0,...,BS3: beam splitters;
phase shifter: aluminum foil;
neutrons that are transmitted by BS1 or BS2 leave the interferometer
and do not contribute to the interference signal.
Detectors count the number of neutrons in the O- and H-beam.
}%
\label{MZIexperiment}
\end{center}
\end{figure*}

Feynman's statement that the event-by-event realization
of an interference pattern is the only mystery suggests
that creating interference patterns by a (local and causal) discrete-event process
may be an important step in demystifying this aspect of quantum phenomena.
Neutron interferometry is a close-to-ideal experimental technique
to address this issue~\cite{RAUC00}.
The basic device used in the neutron interferometry experiments
which are covered in this paper is a Laue-type interferometer~\cite{RAUC74a,RAUC00,HASE11}.
A large, perfect crystal of silicon is cut as shown in Fig.~\ref{MZIexperiment}.
The crystal plate BS0 acts as a beam splitter: neutrons incident from the left
are transmitted with or without being refracted by this plate.
Neutrons refracted by beam splitters BS1 and BS2
are directed to the third plate (BS3) which also acts as a beam splitter.
Neutrons which are not refracted by beam splitters BS1 and BS2 leave the interferometer.
To observe interference, the crystal planes of the different components
have to be parallel to high accuracy~\cite{RAUC74a}
and the whole device needs to be protected from vibrations~\cite{KROU00}.
All beam splitters are assumed to have the same reflection and transmission coefficients~\cite{RAUC00}.
Neutron detectors can have a very high, almost 100\%, efficiency~\cite{RAUC00}.

Many neutron interferometry experiments show that the intensity in the O- and H-beam, obtained by
counting individual neutrons for a certain amount of time,
exhibit sinusoidal variations as a function of the phase shift $\chi$, a prime characteristic of interference~\cite{RAUC00}.
Feynman's ``mystery'' pops up immediately if one wants to entertain
the idea that only waves can produce interference.

Adopting a wave-packet picture for an individual neutron,
the wave packet first splits in two parts at BS0, then each part splits in two at BS1 and BS2.
Two of the four parts go off to infinity,
the other two parts ``reunite'' at BS3.
At BS3 the merged wave packet splits again in two parts.
Only one of these parts triggers a detector.
It is indeed a mystery how four wave packets can conspire to do such things.
Assuming that only a neutron, not merely a part of it can trigger the nuclear
reaction that causes the detector to ``click'',
on elementary logical grounds, the argument that was just given
rules out a wave-packet picture for the individual neutron
(invoking the wave function collapse only adds to the mystery)
but there is no conflict with the statistical interpretation
of quantum mechanics~\cite{BALL03,NIEU11b}.
As long as we consider descriptions of the statistics
of the experiment with many neutrons, we may still think of
one single ``probability'' wave propagating through the interferometer
and as the statistical interpretation of quantum theory is silent
about single events, there is no conflict with logic either~\cite{RAUC00}.
In this paper, we do not solve the aforementioned mystery but give an affirmative answer to the
question whether it is possible to construct a logically consistent,
cause-and-effect description in terms of discrete-event, particle-like processes
which produce results that agree with those of neutron interferometry experiments
and the quantum theory thereof.

In previous work~\cite{RAED05d,RAED05b,MICH11a,ZHAO08b,MICH10a,JIN10c,JIN10b,JIN10a,MICH12a}
we have demonstrated, using an event-based corpuscular model,
that interference is not necessarily a signature of the presence of waves
of some kind but can also appear as the collective result of particles which at any time do not
directly interact with each other.
In general, the event-based approach takes as a starting point the observation that experiments yield definite results, such as
for example the individual detector clicks that build up an interference pattern.
We call these definite results ``events''.
Instead of trying to fit the existence of these events
in some formal, mathematical theory, in the event-based approach the paradigm is changed
by directly searching for the rules that transform events
into other events and, which by repeated application,
yield frequency distributions of events that
agree with those predicted by classical wave or quantum theory.
Obviously, such rules cannot be derived from quantum theory or,
as a matter of fact, of any theory that is probabilistic in nature
simply because these theories do not entail a procedure (= algorithm)
to produce events themselves.

The paper is structured as follows.
In Section~\ref{DLM}, we specify the event-based model in detail.
Sections~\ref{interferometer}--\ref{time}
present our results for the basic neutron interferometer (see Fig.~\ref{MZIexperiment}),
experiments with stochastic and deterministic absorption,
a Bell inequality test, an experiment that creates entanglement
between the neutron path, spin and energy, and
experiments that are performed in a non-stationary regime.
For reference and to facilitate comparison, for each of the experiments that we discuss
in this paper, we give the results of the quantum theoretical description of these experiments,
adopting the terminology that is commonly used in quantum theory.
In contrast, when we discuss the event-by-event, particle-like models of these experiments,
there is no need to invoke concepts such as probability amplitudes, particle-wave duality etc.
Our conclusions and outlook are given in Section~\ref{summary}.

\section{Event-based model}\label{DLM}

The event-based approach has successfully been
used to perform discrete-event simulations of the single beam splitter and
Mach-Zehnder interferometer experiment
of Grangier {\sl et al.}~\cite{GRAN86} (see Refs.~\cite{RAED05d,RAED05b,MICH11a}),
Wheeler's delayed choice experiment of Jacques {\sl et al.}~\cite{JACQ07}
(see Refs.~\cite{ZHAO08b,MICH10a,MICH11a}),
the quantum eraser experiment of Schwindt {\sl et al.}~\cite{SCHW99} (see Ref.~\cite{JIN10c,MICH11a}),
double-slit and two-beam single-photon interference experiments and the single-photon interference experiment with
a Fresnel biprism of Jacques {\sl et al.}~\cite{JACQ05} (see Ref.~\cite{JIN10b,MICH11a}),
quantum cryptography protocols (see Ref.~\cite{ZHAO08a}),
the Hanbury Brown-Twiss experiment of Agafonov {\sl et al.}~\cite{AGAF08} (see Ref.~\cite{JIN10a,MICH11a}),
universal quantum computation (see Ref.~\cite{RAED05c,MICH05}),
Einstein-Podolsky-Rosen-Bohm-type experiments of Aspect {\sl et al.}~\cite{ASPE82a,ASPE82b}
and of Weihs {\sl et al.}~\cite{WEIH98} (see Refs.~\cite{RAED06c,RAED07a,RAED07b,RAED07c,RAED07d,ZHAO08,MICH11a}),
and the propagation of electromagnetic plane waves through homogeneous thin films and stratified media (see Ref.~\cite{TRIE11,MICH11a}).
An extensive review of the simulation method and its applications is given in Ref.~\cite{MICH11a}.
Proposals for single-particle experiments to test specific aspects of the event-based approach are discussed in Refs.~\cite{JIN10b,MICH12a}.
For many different optics experiments,
the event-based corpuscular model reproduces the probability distributions of quantum theory
or results of Maxwell's wave theory by assuming that photons have a particle character only.

The event-based corpuscular model is free of paradoxes that result from the assumption that photons exhibit a dual,
wave-particle behavior,
and as we demonstrate in this paper, the same holds for neutrons as well.
A crucial property of the event-based corpuscular models is that they reproduce various ``wave results''
observed in different experiments without any change
to algorithms modeling the particles and components (e.g. beam splitters)~\cite{MICH11a}.
Although the event-based algorithms can be given an interpretation of a realistic cause-and-effect description that
is free of logical difficulties, in the present stage of development
it is difficult to decide whether or not such algorithms or modifications of them are realized by Nature.
Only new, dedicated experiments may teach us more about this intriguing question.

\subsection{Definition of messenger and message}

A neutron is regarded as a messenger, carrying a message.
As in our earlier event-based models for quantum optics experiments~\cite{MICH11a},
we represent a message by the two-dimensional complex-valued unit vector
\begin{equation}
\mathbf{y}=
\left(\begin{array}{c}
        e^{i\psi^{(1)}} \cos(\theta/2)\\
        e^{i\psi^{(2)}} \sin(\theta/2)
\end{array}\right)
.
\label{mess2}
\end{equation}
As is often the case, it is convenient though by no means essential to work with complex-valued vectors.
The message Eq.~(\ref{mess2}) encodes the time of flight and the magnetic moment of the neutron.

In a pictorial manner, the neutron carries with it a clock, the hand of which
rotates with angular frequency $\nu$ (to be discussed later).
The clock may be used by event-based processors, mimicking the interaction of neutrons with materials,
to determine the neutron's time of flight.
Similarly, if we think of the neutron as a tiny classical magnet spinning
around the direction $\mathbf{m}=(\cos\phi\sin\theta,\sin\phi\sin\theta,\cos\theta)$, relative
to a fixed frame of reference defined by a magnetic field,
then, the two angles $\phi$ and $\theta$ suffice to specify the magnetic moment.

According to Eq.~(\ref{mess2}), within the present model, it is postulated  that the internal state of the neutron
is fully determined by the three angles $\psi^{(1)}$, $\psi^{(2)}$, and $\theta$
and by rules, to be specified, by which these angles change as the neutron
moves through space.
In Eq.~(\ref{mess2}), we have introduced three angles to characterize the message.
The difference $\phi=\psi^{(1)}-\psi^{(2)}$ and $\theta$ suffice to represent the magnetic moment
and the third degree of freedom is used to account for the time of flight of the neutron.

At this stage of the development, it is not clear whether the model of the messenger that we describe here
is sufficient to explain all possible neutron interferometry experiments that might be carried out
but to explain the neutron interferometry experiments
which are covered in this paper, it cannot be simplified further.

As the messenger moves for a time $T$, it is postulated that the message changes according to the rule
\begin{equation}
\mathbf{y}\leftarrow e^{i\nu T}\mathbf{y}
,
\label{mess3}
\end{equation}
where $T$ is the time of flight, relative to the time of creation of the messenger,
and $\nu$ is an angular frequency.
A monochromatic beam of incident neutrons is assumed to consist of
neutrons that all have the same value of $\nu$~\cite{RAUC00}.
Put differently, of all the neutrons created in the fission process,
the purpose of the monochromator is to select those neutrons that share
the same characteristics: the velocity and direction in a classical mechanical picture~\cite{RAUC00}
and the angular frequency $\nu$ and direction in the event-based picture.
The direction merely serves to send the selected neutrons to the interferometer.

In the event-based picture, messengers can travel along a single path only.
As they travel through the interferometer (one at a time) and are detected
by one of the detectors, their times of flight may be different from messenger to messenger, depending
on which path they followed and the delay they experienced in the material that acts as a phase shifter.
Still within the event-based picture, the experimental fact
that the measured intensity depends on the position of the phase shifter
is a direct proof that the messenger conveys its time of flight to the processors.
Hence it must have some kind of internal clock.

A very plausible choice would be to relate $\nu$ to the energy $E$ of the neutron,
that is we could make the hypothesis that $\nu\propto E/h$ where Planck's constant
appears as a scale factor to render $\nu T$ dimensionless.
However, in this paper, the emphasis is on demonstrating that
a particle-only model can reproduce the interference phenomena observed in
neutron interferometry experiments and to simplify matters,
we only consider idealized experiments with monochromatic beams of neutrons.
In this case, the actual value of $\nu$ does not affect the detector counts.
Event-based simulations of experiments in which the actual value(s) of $\nu$ are
important, e.g. experiments which involve gravitation~\cite{OVER74,COLE75,RAUC00,JENK11},
are left for future research.

In the presence of a magnetic field, a magnetic moment rotates about
the direction of the magnetic field according to the standard, classical equation of motion.
In terms of the message, this corresponds to a rotation of $\mathbf{y}$
about the same direction.
As Eq.~(\ref{mess2}) suggests, the magnetic moment is represented through
the well-known Bloch-sphere representation of a spin-1/2 particle~\cite{BALL03}.
Exploiting the relation between rotations in three-dimensional space and rotations in spin-1/2 Hilbert space,
in the presence of a magnetic field, the message changes according to the rule
\begin{equation}
\mathbf{y}\leftarrow e^{i(\sigma^xB_x+\sigma^yB_y+\sigma^zB_z)}\mathbf{y}
,
\label{mess4}
\end{equation}
where $\sigma^x$, $\sigma^y$, and $\sigma^z$ are the Pauli spin-matrices
and $\mathbf{B}=(B_x,B_y,B_z)$ denotes the magnetic field vector.
Although Eq.~(\ref{mess4}) is reminiscent of the rotation operator of a spin-1/2 quantum object,
in the present context Eq.~(\ref{mess4}) is just a convenient construct to implement rotations in
three-dimensional space.

\subsection{Particle source}

The source creates messengers and initializes the message.
In order to demonstrate that the class of models which we consider can produce
interference without solving wave equations,
we explicitly exclude the possibility that at any time there is more than
one messenger passing through the interferometer, an assumption which is often made
in the discussion of neutron interferometry experiments~\cite{UNNE90}.
In the simulation, it is trivial to realize this condition:
except for the first particle, the source creates a new particle
only after the previous particle has been detected.
It is also straightforward to let the source produce particles with specific properties.
For instance, a fully coherent spin-polarized beam is simulated by
generating messengers with the message given by Eq.~(\ref{mess2})
where $\psi^{(1)}$, $\psi^{(2)}$, and $\theta$ are the same for all messages.
Throughout this paper, the total number of particles generated by the source is denoted by $N$.

\subsection{Beam splitter}\label{BS}

\begin{figure}[t]
\begin{center}
\includegraphics[width=12cm]{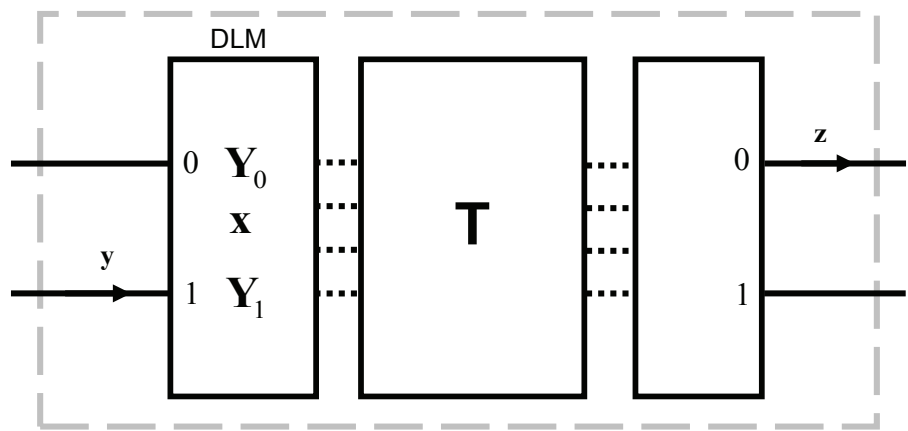}
\caption{Diagram of a DLM-based processing unit that performs an event-based simulation of
the beam splitters in the neutron interferometer (see Fig.~\ref{MZIexperiment}).
The processing unit consists of three stages: an input stage (DLM), a transformation stage and an output stage.
The solid lines represent the input and output ports of the device.
The presence of a message $\mathbf {y}$ (incident neutron) is indicated by an arrow on the corresponding port line.
The DLM has storage for two real numbers ($\mathbf{x}$) and two complex vectors $\mathbf{Y}_0$ and $\mathbf{Y}_1$
that are updated according to the rules Eqs.~(\ref{int20}) and (\ref{int21}), respectively.
This data is combined to yield a 4-dimensional complex-valued vector which, after transformation by
a matrix $\mathbf{T}$, is fed into the output stage which decides
through which port the (modified) message $\mathbf {z}$ leaves the device.
The dashed lines indicate the data flow within the unit.
}
\label{figmachine0}
\end{center}
\end{figure}

In Fig.~\ref{figmachine0}, we show the diagram of the event-based processor that
simulates the operation of a beam splitter.
This processor has three stages.
The input stage consists of a so-called deterministic learning machine (DLM)~\cite{RAED05b,RAED05d,MICH11a}.
This machine is capable of learning, on the basis of the individual events,
about the relative frequencies of messengers arriving on ports $0$ and $1$.
In neutron interferometry experiments, it is assumed that at any time,
at most one neutron passes through the interferometer~\cite{RAUC74a,RAUC00}.
In the event-based approach, this assumption implies
that the DLM receives a message on either input port $0$ or $1$, never on both ports simultaneously.

The arrival of a messenger at port 0 or 1 is represented by the vectors ${\bf v}=(1,0)$ or ${\bf v}=(0,1)$,
respectively.
A DLM that is capable of performing the desired task
has an internal vector $\mathbf{x}=( x_{0},x_{1})$,
where $x_{0}+x_{1}\le1$ and $x_{k}\geq 0$ for all $k=0,1$.
In addition to the internal vector $\mathbf{x}$,
the DLM should have two sets of two registers
$\mathbf{Y}_{k}=(Y_{k,1},Y_{k,2})$
to store the last message $\mathbf{y}$ that arrived at port $k$.
Thus, the DLM has storage for exactly 10 real numbers.

Upon receiving a messenger at input port $k$, the DLM performs the following steps:
it copies the elements of message $\mathbf{y}$ in its internal register $\mathbf{Y}_k$
\begin{equation}
\mathbf{Y}_k\leftarrow\mathbf{y}
\label{int21}
\end{equation}
while leaving $\mathbf{Y}_{1-k}$ unchanged,
and updates its internal vector $\mathbf{x}$ according to
\begin{equation}
\mathbf{x}\leftarrow\gamma \mathbf{x}+( 1-\gamma ) \mathbf{v}
.
\label{int20}
\end{equation}
It is easy to see that $x_0+x_1\le1$ at all times.
Each time a messenger arrives at one of the input ports,
the DLM updates the values of the internal vector $\mathbf{x}$
and overwrites the values in the registers $\mathbf{Y}_{k}$.
Thus, the machine can only store data of two messengers, not of all of them.

The parameter $0\le\gamma<1$ affects the number of events the machine needs to
adapt to a new situation, that is when the ratio of particles on paths 0 and 1 changes.
By reducing $\gamma$, the number of events needed to adapt decreases but the accuracy with which the machine
reproduces the ratio also decreases. In the limit that $\gamma=0$, the machine learns nothing:
it simply echoes the last message that it received~\cite{RAED05b,RAED05d}.
If $\gamma\rightarrow1^-$, the machine learns slowly and reproduces accurately the ratio
of particles that enter via port 0 and 1.
It is in this case that the machine can be used to reproduce, event-by-event, the
interference patterns that are characteristic of quantum phenomena~\cite{RAED05b,RAED05d,MICH11a}.

\begin{figure}[t]
\begin{center}
\includegraphics[width=12cm]{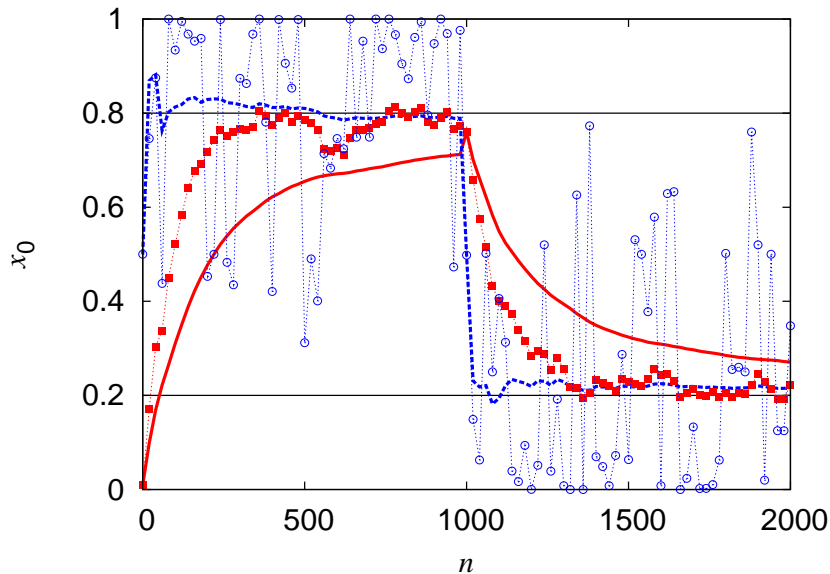}
\caption{Simulation data of the internal variable $x_0$ as a function of the number of received
input events $n$, generated by Eq.~(\ref{int20}). Initially $x_0=0$.
For $n=0,\ldots,999$, the input is either $\mathbf{v}=(1,0)$ ($\mathbf{v}=(0,1)$) with probability $0.8$ ($0.2$).
For $n=1000,\ldots,2000$, the input is either $\mathbf{v}=(1,0)$ ($\mathbf{v}=(0,1)$) with probability $0.2$ ($0.8$).
The horizontal lines represent the probabilities $0.8$ and $0.2$
The data of $x_0$ is shown as markers connected by thin lines.
The running average of $x_0$ is shown as a thick line.
At $n=1001$, the value of the running average is set equal to the current value of $x_0$.
Solid squares and solid line: $\gamma=0.99$;
Open circles and dashed line: $\gamma=0.5$.
The data is plotted for every 20 input events.
}
\label{dlmdemo}
\end{center}
\end{figure}

For later applications, it may be useful to have some insight into the dynamics of this DLM.
In Fig.~\ref{dlmdemo}, we show some representative results obtained by executing the rule Eq.~(\ref{int20}).
From Fig.~\ref{dlmdemo}, it is clear that all the features that we have  discussed are present in the data.
As $\gamma=0.99$ is close to one, the processor learns slowly.
It takes several hundreds of input events before $x_0$ fluctuates around the probability $0.8$ for a $(1,0)$ input event.
If we change the latter from 0.8 to 0.2, the DLM reacts immediately but again it takes a few hundred steps
to reach the stationary state that corresponds to the new input sequence.
The relatively slow pace with which the DLM responds to a change of the input sequence
has a significant impact on the running average, represented by the thick solid line.
For the number of events shown, for  $\gamma=0.99$, the running average does not come close
to its asymptotic value (for 20000 input events it does, data not shown).
For $\gamma=0.5$, we see that the DLM responds very fast but this at the cost of large fluctuations of $x_0$.

For applications to the event-based simulation of quantum phenomena, the value of the running average is of no importance but
the fluctuations of $x_0$ are. We will see later that large fluctuations reduce the visibility of the interference signal.
The number of events, required to establish the stationary state, becomes an important issue
for simulating experiments in which the conditions rapidly change with time (see Section~\ref{shutter}).
Otherwise it is not an issue.
Summarizing: the parameter $\gamma$ determines the ``quality'' of the event-by-event model of the interferometer,
the ideal interferometer
corresponding to $\gamma\rightarrow1^-$.

Returning to the diagram of the processor,
the second stage accepts a message from the input stage and transforms it into a new message.
From the description of the DLM, it is clear that
the internal registers $\mathbf{Y}_0$ and $\mathbf{Y}_1$
contain the last message that arrived on input port 0 and 1 respectively.
First, this data is combined with the data of the internal vector
$\mathbf{x}$, the components of which converge (after many events
have been processed) to the relative frequencies with which
the messengers arrive on port 0 and 1, respectively.
The output message generated by the transformation stage is
\begin{equation}
\left(\begin{array}{cc}
        {Z}_{0,1}\\
        {Z}_{1,1}\\
        {Z}_{0,2}\\
        {Z}_{1,2}
\end{array}\right)
=
\left(\begin{array}{cccc}
        \phantom{i}\sqrt{T}&i\sqrt{R}&0&0\\
        i\sqrt{R}&\phantom{i}\sqrt{T}&0&0\\
        0&0&\phantom{i}\sqrt{T}&i\sqrt{R}\\
        0&0&i\sqrt{R}&\phantom{i}\sqrt{T}
\end{array}\right)
\left(\begin{array}{cccc}
        x_0^{1/2}&0&0&0\\
        0&x_1^{1/2}&0&0\\
        0&0&x_0^{1/2}&0\\
        0&0&0&x_1^{1/2}
\end{array}\right)
\left(\begin{array}{c}
        {Y}_{0,1}^{\phantom{\prime}}\\
        {Y}_{1,1}^{\phantom{\prime}}\\
        {Y}_{0,2}^{\phantom{\prime}}\\
        {Y}_{1,2}^{\phantom{\prime}}
\end{array}\right)
,
\label{trans0}
\end{equation}
where the reflection $R$ and transmission $T=1-R$ are real numbers that are
considered to be parameters, to be determined from experiment.
Note that in contrast to optics~\cite{MICH11a} where S- and P-polarized waves may behave differently upon reflection/transmission~\cite{BORN64},
in the case of neutrons, the first matrix in Eq.~(\ref{trans0}) (reading from left to right)
treats the first and second pair of the four-dimensional vector on equal footing,
in concert with the quantum theoretical treatment in section~\ref{qt1}.
Further note that as $x_0+x_1\le1$ at all times and $\Vert\mathbf{Y}_{0}\Vert=\Vert\mathbf{Y}_{1}\Vert=1$,
we have $|{Z}_{0,1}|^2 +|{Z}_{0,2}|^2 +|{Z}_{1,1}|^2 +|{Z}_{1,2}|^2=1$.

The output stage uses the data provided by the transformation stage
to decide through which of the two ports a messenger (representing a neutron) will be sent.
The rule is very simple.
We compute $z=|{Z}_{1,1}|^2+|{Z}_{1,2}|^2$ and select the output port $k'$ by the rule
\begin{equation}
k' = \Theta(z - \RN)
,
\label{out0}
\end{equation}
where $\Theta(.)$ is the unit step function
and $0\leq \RN <1$ is a uniform pseudo-random number (which changes with each messenger processed).
From a simulation point of view, there is nothing special about using pseudo-random numbers.
On a digital computer, pseudo-random numbers are generated by deterministic processes and therefore
the pseudo-random number generator may be replaced by any algorithm
that selects the output port in a systematic, uniform manner~\cite{RAED05b,MICH11a},
as long as the zero's and one's occur with a ratio determined by $z$.
In fact, we use pseudo-random numbers to mimic the apparent unpredictability
of the experimental data only.

The messenger leaves through either port $k'=0$ or port $k'=1$ carrying the message
\begin{equation}
\mathbf{z} =
\frac{1}{\sqrt{|{Z}_{k',1}|^2+|{Z}_{k',2}|^2}}
\left(\begin{array}{c}
        {Z}_{k',1}\\
        {Z}_{k',2}
\end{array}\right)
,
\label{out1}
\end{equation}
which, for internal consistency and modularity of the event-based approach,
is also a unit vector.

\subsection{Detector}

In the simulation model, we simply count all neutrons that leave the apparatus through the O- and H-beam.
In other words, we assume that the detectors have 100\% detection efficiency.
Note that real neutron detectors can have efficiencies of 99\% and more~\cite{KROU00}.

\begin{figure*}[t]
\begin{center}
\includegraphics[width=12cm ]{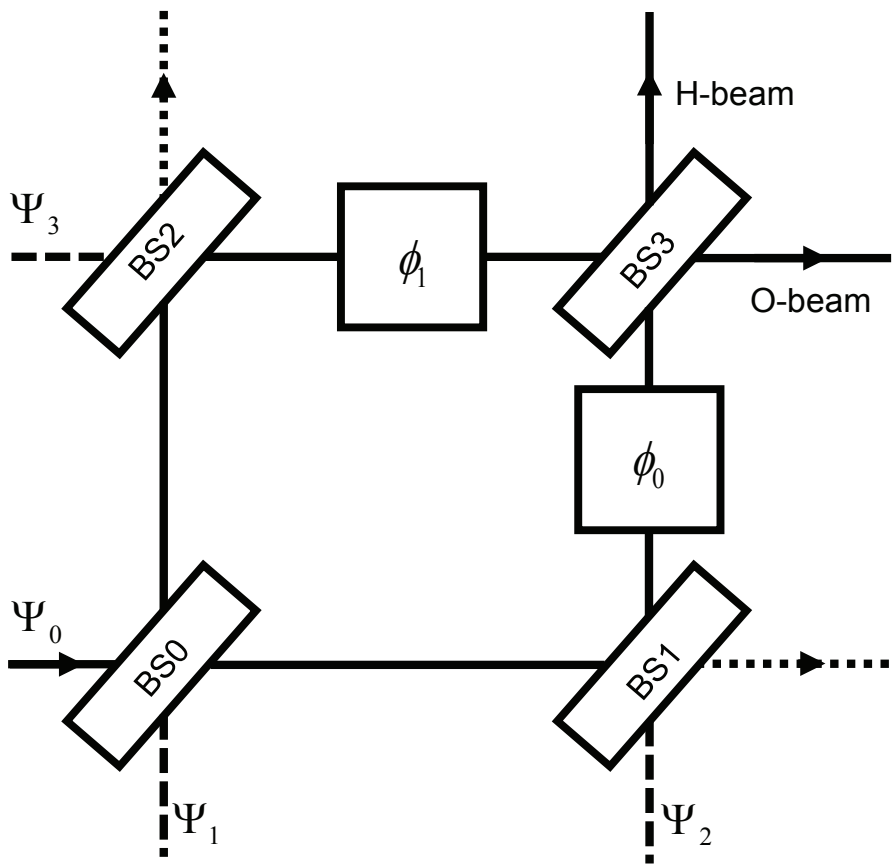}
\caption{%
Diagram of the interferometer shown in Fig.~\ref{MZIexperiment}.
BS0,...,BS3: beam splitters;
$\phi_0$ and $\phi_1$: phase shifters.
Detectors count all particles that leave the interferometer via the O- and H-beam.
In experiment and in the event-based simulation,
neutrons enter the interferometer via the path labeled $\Psi_0$ only.
The paths labeled $\Psi_1$, $\Psi_2$, and $\Psi_3$
are used in the quantum theoretical treatment only (see text).
Particles leaving the interferometer via the dotted lines are not counted.
}%
\label{qt-mzi}
\end{center}
\end{figure*}

\subsection{What makes it work?}

Anticipating that the event-based processor described in this section
will perform as expected, that is, produce the expected interference patterns,
it may be useful to have a deeper understanding of how it can be that these
patterns appear without solving a wave problem.

Let us consider BS3 in Fig.~\ref{MZIexperiment},
the beam splitter at which, in a wave picture,
the two beams join to produce interference.
The event-based processor simulating a beam splitter requires two pieces
of information to send out particles such that
their distribution matches the wave-mechanical
description of the beam splitter.
First, it needs an estimate of the ratio
of particle currents in the O- and H-beam, respectively.
Second, it needs to have information about the
time of flight along the two different paths.

The first piece of information is provided for by
the internal vector $\mathbf{x}$.
As explained above, through the update rule Eq.~(\ref{int20}),
for a stationary sequence of input events,
$\mathbf{x}=(x_0,x_1)$ converges to the average
of the number of events on input ports 0 and 1, respectively.
Thus, the intensities of the waves in the two input beams
are encoded in the vector $\mathbf{x}$.
Note that this information is accurate only if
the sequence of input events is stationary.

After one neutron arrived at port 0 and another
one arrived at port 1, the second piece of information is always
available in the registers $\mathbf{Y}_0$ and $\mathbf{Y}_1$.
This information plays the role of the phase
of the waves in the two input beams.

It is now easy to see that all the information (intensity and phase)
is available to compute the probability for sending
out particles according to the distribution
that we know from wave mechanics.
Indeed, in the stationary state, Eq.~(\ref{trans0})
is identical to the transformation of the wave amplitudes
which we know from wave theory of a beam splitter~\cite{RAUC00,BORN64}.

The idea that the event-based model of a beam splitter
has some memory and a learning capability may seem strange enough
to reject the model at first sight.
However, applying the same logic to for instance Maxwell's
theory of electrodynamics, one should reject this model as well.
Indeed, the interaction of the electromagnetic wave
and a material invariably takes a form that involves memory.
This can be seen as follows.
In Maxwell's theory, for electromagnetic radiation
with frequency $\omega$, the (linear part of the) interaction
of the electric field $\mathbf{E}(\omega)$ and a material
takes the form $\mathbf{P}(\omega)=\eta(\omega)\mathbf{E}(\omega)$
where $\mathbf{P}(\omega)$ and $\eta(\omega)$ are the
polarization and dielectric susceptibility of the material, respectively~\cite{BORN64}.
Transforming this relation to the time domain and assuming that $\mathbf{E}(t=0)=\mathbf{P}(t=0)=0$
yields~\cite{TAFL05}
\begin{equation}
\mathbf{P}(t)=\int_0^t \eta(t-u)\mathbf{E}(u)\, du
,
\label{mem0}
\end{equation}
where the memory kernel $\eta(t)$ is the Fourier transform of $\eta(\omega)$.
Clearly, Eq.~(\ref{mem0}) shows that the response of the polarization vector
to the electric field involves memory.

It is instructive to make the analogy with the update rule Eq.~(\ref{int20}) more explicit.
Assume that $\mathbf{x}_k$ and $\mathbf{v}_k$ are
the values of time-dependent vectors $\mathbf{x}(t)$ and $\mathbf{v}(t)$ sampled at regular
time intervals $\tau$.
If $\mathbf{x}(t)$ allows a Taylor series expansion, we may write
$\mathbf{x}_k=\mathbf{x}(\tau k)$,
$\mathbf{x}_{k-1}=\mathbf{x}(\tau k)-\tau d\mathbf{x}(t)/dt|_{t=\tau k}+{\cal O}(\tau^2)$ such
that the update rule Eq.~(\ref{int20}) can be expressed as
\begin{eqnarray}
\frac{d \mathbf{x}(t)}{dt}=-\frac{1-\gamma}{\tau\gamma}\mathbf{x}(t)+\frac{1-\gamma}{\tau\gamma}\mathbf{v}(t)
.
\label{dlms7}
\end{eqnarray}
In order that Eq.~(\ref{dlms7}) makes sense for $\tau\rightarrow0$, we must have
$\lim_{\tau\rightarrow0}(1-\gamma)/\tau\gamma=\Gamma$.
This requirement is trivially satisfied by putting $\gamma=1/(1+\tau\Gamma)$.
Then Eq.~(\ref{dlms7}) takes the form of the first-order linear differential equation
\begin{eqnarray}
\frac{d \mathbf{x}(t)}{dt}=-\Gamma\mathbf{x}(t)+\Gamma\mathbf{v}(t)
.
\label{dlms8}
\end{eqnarray}
Assuming $\mathbf{x}(0)=0$, the formal solution of Eq.~(\ref{dlms8}) reads
\begin{eqnarray}
\mathbf{x}(t)=\Gamma\int_0^t e^{-u\Gamma}\mathbf{v}(t-u)du
,
\label{dlms9}
\end{eqnarray}
which has the same structure as Eq.~(\ref{mem0}).
From the derivation of Eq.~(\ref{dlms8}), it follows that
if we interpret $\tau$ as the time interval between two successive messages
and let $\tau$ approach zero, then $\gamma=1/(1+\tau\Gamma)$ approaches one and the DLM defined by the update rule
Eq.~(\ref{int20}) ``solves'' the differential equation Eq.~(\ref{dlms8}).
Therefore, we may view Eq.~(\ref{dlms8}) as a course-grained, continuum approximation to the
event-by-event process defined by Eq.~(\ref{int20}).

Summarizing, the general idea that objects retain some ``memory'' about their interaction with
external agents (particles, fields,...) is not only common but even essential to some of
the most successful theories of physical phenomena
and can therefore not be used as an argument to dismiss a particular class of models.
Furthermore, it is worth noting that Eq.~(\ref{int20}) is not the only update rule
which yields an event-based model that reproduces the averages predicted by quantum theory~\cite{RAED05b,MICH11a}.
In other words, there is nothing ``unique'' to Eq.~(\ref{int20}).
Whether an event-based model accounts for what is actually happening on the level of single events
can only be decided by experiments that address this specific question.

\section{Neutron interferometer}\label{interferometer}

A detailed wave-mechanical description of the diffraction of neutrons
by the perfect silicon plate and the complete neutron interferometer is
given in Ref.~\cite{RAUC00}.
In this paper, to simplify matters without giving in on the fundamental issues,
we adopt an effective model for the scattering process of the neutron and the plate.
We assume that the neutrons are monochromatic and
satisfy the Bragg condition for scattering by the silicon plate~\cite{RAUC00}.
This is not an essential simplification.
In the theory of neutron interferometry, it is customary
to compute the incoherent average over slight deviations from the exact Bragg condition
and neutron energy~\cite{RAUC00}
and the same can be done in the event-based approach as well (see Section~\ref{shutter}).
Thus, we will characterize the beam splitters BS0,...,BS3  by
effective reflection and transmission coefficients $r$ and $t$, respectively.

Once it has been established that the event-based approach reproduces
the results of wave theory, a ray-tracing scheme such as the one outlined
in Ref.~\cite{RAUC00} can be combined with the event-based
processors to yield a more complete description of how the individual
neutrons propagate through the interferometer and produce interference.
We leave this technically challenging topic for future research.

\subsection{Quantum theory}\label{qt1}

A detailed quantum mechanical treatment of the interferometer depicted in Fig.~\ref{MZIexperiment}
is given in Ref.~\cite{RAUC74b}.
Assuming that the incident wave satisfies the Bragg condition for scattering by the
first crystal plate (BS0), the Laue-type interferometer acts as a two-path
interferometer~\cite{HORN86}.
The two-path interferometer may be represented by a more abstract, theoretical model,
the diagram of which is shown in Fig.~\ref{qt-mzi}.
This diagram is similar to the one of the Mach-Zehnder interferometer for light~\cite{BORN64},
except that the latter has mirrors instead of beam splitters BS1 and BS2.

Quantum theory describes the statistics of the interferometry experiment depicted in Figs.~\ref{MZIexperiment}
and \ref{qt-mzi} in terms of the state vector
\begin{equation}
|\Psi\rangle=
\left( \Psi_{0\uparrow}, \Psi_{0\downarrow}, \Psi_{1\uparrow}, \Psi_{1\downarrow}
       \Psi_{2\uparrow}, \Psi_{2\downarrow}, \Psi_{3\uparrow}, \Psi_{3\downarrow}
\right)^T
,
\label{app0}
\end{equation}
where the components of this vector represent the complex-valued amplitudes of the wave function.
The first subscript labels the pathway and
the second subscript denotes the direction of the magnetic moment relative
to some B-field, the direction of which becomes relevant if the experimental outcome depends
on the magnetic moment of the neutron (see later).
This is not the case for the experiment shown in Fig.~\ref{MZIexperiment},
hence there is no need to dwell on this aspect any further.
As usual, the state vector is assumed to be normalized, meaning that $\langle\Psi|\Psi\rangle=1$.
Note that in the abstract representations of the experiments, such as in Fig.~\ref{qt-mzi} for example,
we use the notation $\Psi_j=(\Psi_{j\uparrow}, \Psi_{j\downarrow})$ for $j=0,\ldots,3$.

As the state vector propagates through the interferometer, it changes according to
\begin{eqnarray}
|\Psi'\rangle
&=&
\left(\begin{array}{cc}
        \phantom{-}\T^\ast &\R\\
        -\R^\ast & \T
\end{array}\right)_{5,7}
\left(\begin{array}{cc}
        \phantom{-}\T^\ast &\R \\
        -\R^\ast & \T
\end{array}\right)_{4,6}
\left(\begin{array}{cc}
         \sqrt{b}e^{i\phi_1}&0 \\
         0 & \sqrt{b}e^{i\phi_1}
\end{array}\right)_{6,7}
\left(\begin{array}{cc}
         \sqrt{a}e^{i\phi_0}&0 \\
         0 & \sqrt{a}e^{i\phi_0}
\end{array}\right)_{4,5}
\nonumber \\ &&\times
\left(\begin{array}{cc}
        \phantom{-}\T^\ast &\R\\
        -\R^\ast & \T
\end{array}\right)_{3,7}
\left(\begin{array}{cc}
        \phantom{-}\T^\ast &\R \\
        -\R^\ast & \T
\end{array}\right)_{2,6}
\left(\begin{array}{cc}
        \T & -\R^\ast \\
        \R & \phantom{-}\T^\ast
\end{array}\right)_{1,5}
\left(\begin{array}{cc}
        \T & -\R^\ast \\
        \R & \phantom{-}\T^\ast
\end{array}\right)_{0,4}
\left(\begin{array}{cc}
        \T & -\R^\ast \\
        \R & \phantom{-}\T^\ast
\end{array}\right)_{1,3}
\left(\begin{array}{cc}
        \T & -\R^\ast \\
        \R & \phantom{-}\T^\ast
\end{array}\right)_{0,2}
|\Psi\rangle
,
\label{app1}
\end{eqnarray}
where $\T$ and $\R$ denote the transmission and reflection coefficients, respectively,
and the subscripts $i,j$ refer to the pair of elements of the
eight-dimensional vector on which the matrix acts.
Conservation of probability demands that $|\T|^2 + |\R|^2 =1$.
For a later application, we have included in Eq.~(\ref{app1}), a path-dependent absorption
parameterized by the coefficients $a$ ($0\le a\le1$)
and $b$ ($0\le b\le1$).

In neutron interferometry experiments, particles enter the interferometer via
the path corresponding to the amplitude $\Psi_0$ only (see Fig.~\ref{qt-mzi}),
meaning that $|\Psi\rangle=(1,0,0,0,0,0,0,0)$.
The probabilities to observe a particle leaving the interferometer
in the H- and O-beam are then given by
\begin{eqnarray}
p_\mathrm{H}=|\Psi'_{2\uparrow}|^2+|\Psi'_{2\downarrow}|^2&=&R\left( aT^2+bR^2 -2 RT\sqrt{ab}\cos\chi \right)
,
\label{app2z}
\\ 
p_\mathrm{O}=|\Psi'_{3\uparrow}|^2+|\Psi'_{3\downarrow}|^2&=&R^2T\left( a+b+ 2\sqrt{ab}\cos\chi \right)
,
\label{app2}
\end{eqnarray}
where $\chi=\phi_1-\phi_0$ is the relative phase shift, $R=|\R|^2$ and $T=|\T|^2=1-R$.
Note that $p_\mathrm{H}$ and $p_\mathrm{O}$ do not depend on the imaginary part of $\T$ or $\R$,
leaving only one free model parameter (e.g. $R$).
In the case of a 50-50 beam splitter ($T=R=1/2$) and zero absorption ($a=b=1$),
Eqs.~(\ref{app2z}) and (\ref{app2}) reduce to the familiar expressions
$p_\mathrm{H}=(1/2)\sin^2\chi/2$ and $p_\mathrm{O}=(1/2)\cos^2\chi/2$, respectively.
The extra factor two is due to the fact
that one half of all incoming neutrons, that is the neutrons that are
transmitted by BS1 or BS2 (see Fig.~\ref{MZIexperiment}), leave the interferometer without being counted.

The expression Eq.~(\ref{app2}) shows that the normalized (to the maximum value) O-beam intensity
does not depend on the value of the reflection $R$.
Furthermore, it follows from Eq.~(\ref{app2}) that the visibility of the O-beam is given by
\begin{eqnarray}
{V}(a,b)&\equiv&
\frac{\max_{\chi} p_\mathrm{O} - \min_{\chi} p_\mathrm{O}}{\max_{\chi} p_\mathrm{O} + \min_{\chi} p_\mathrm{O}}
=\frac{2\sqrt{ab}}{a+b}
,
\label{app2a}
\end{eqnarray}
and that the modulation amplitude of the interference fringes is given by $(a+b)V(a,b)/2=\sqrt{ab}$.

\begin{figure*}[t]
\begin{center}
\includegraphics[width=8cm ]{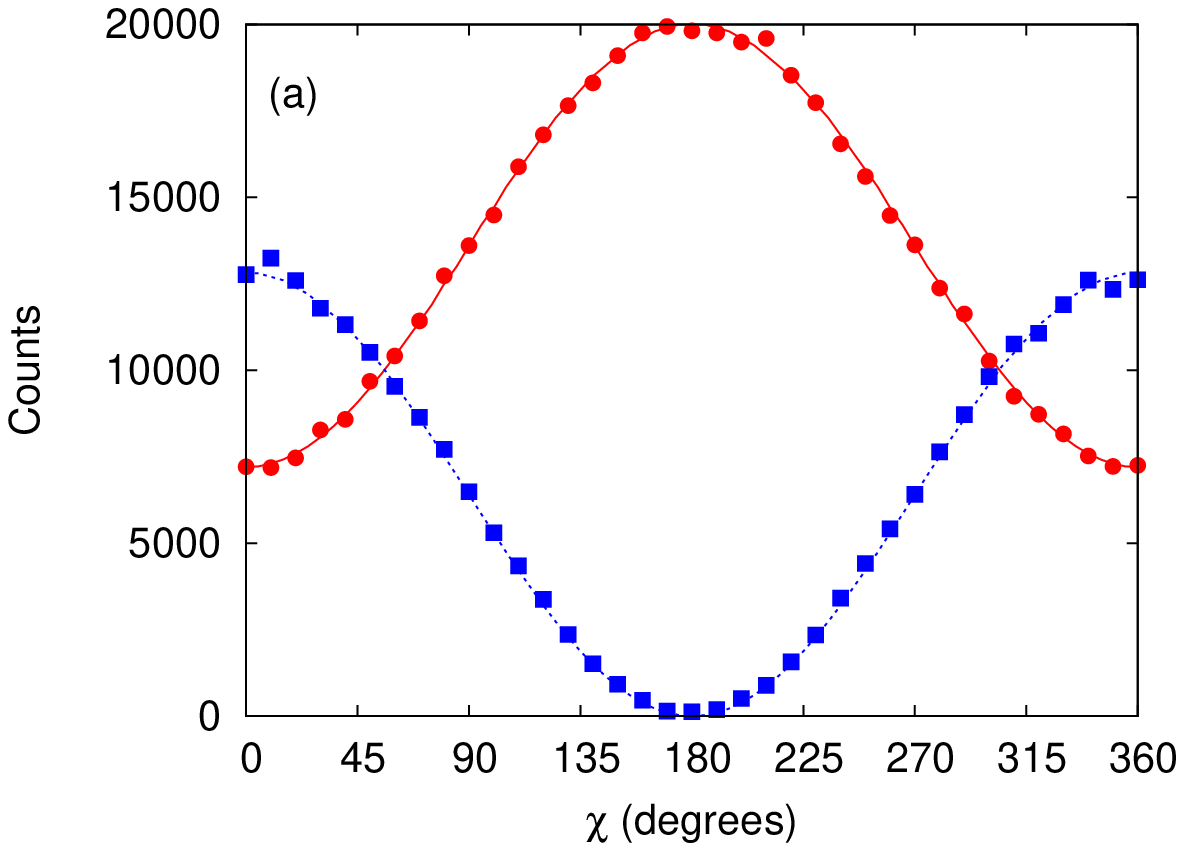}
\includegraphics[width=8cm ]{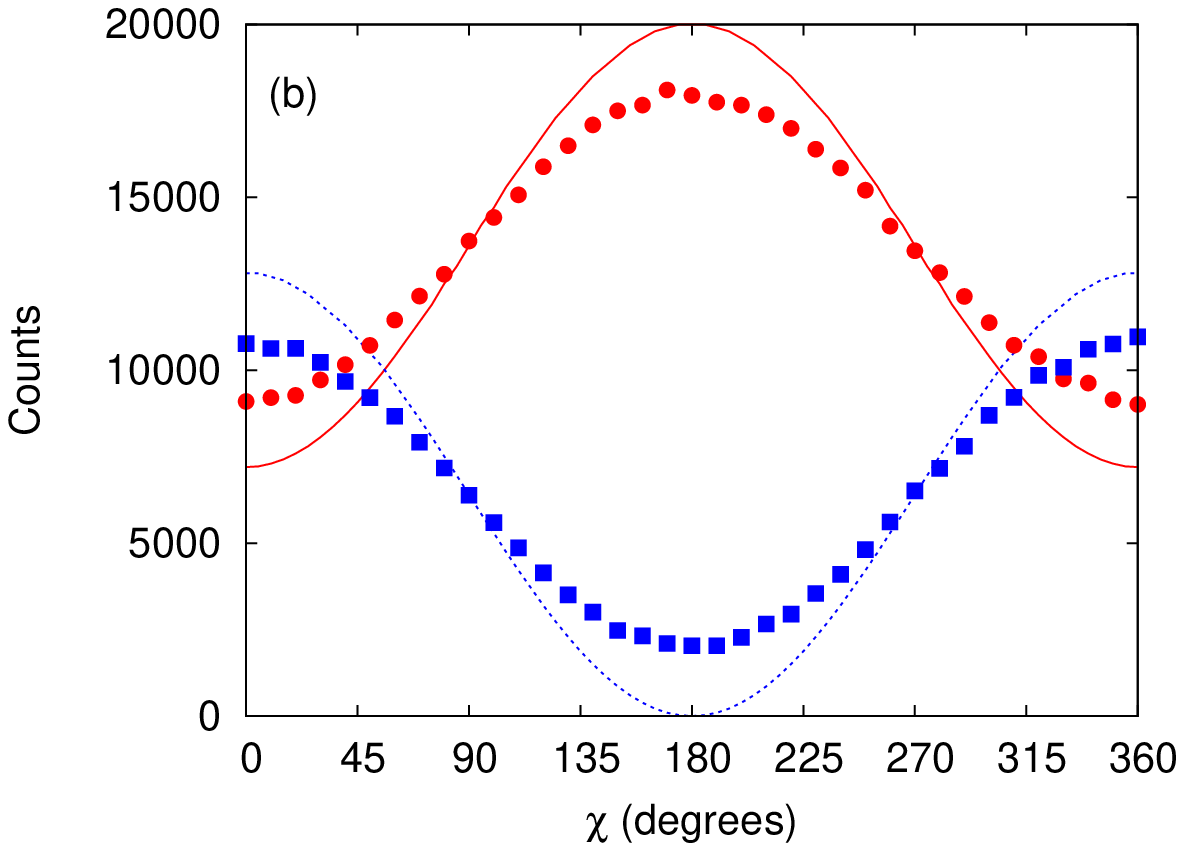}
\includegraphics[width=8cm ]{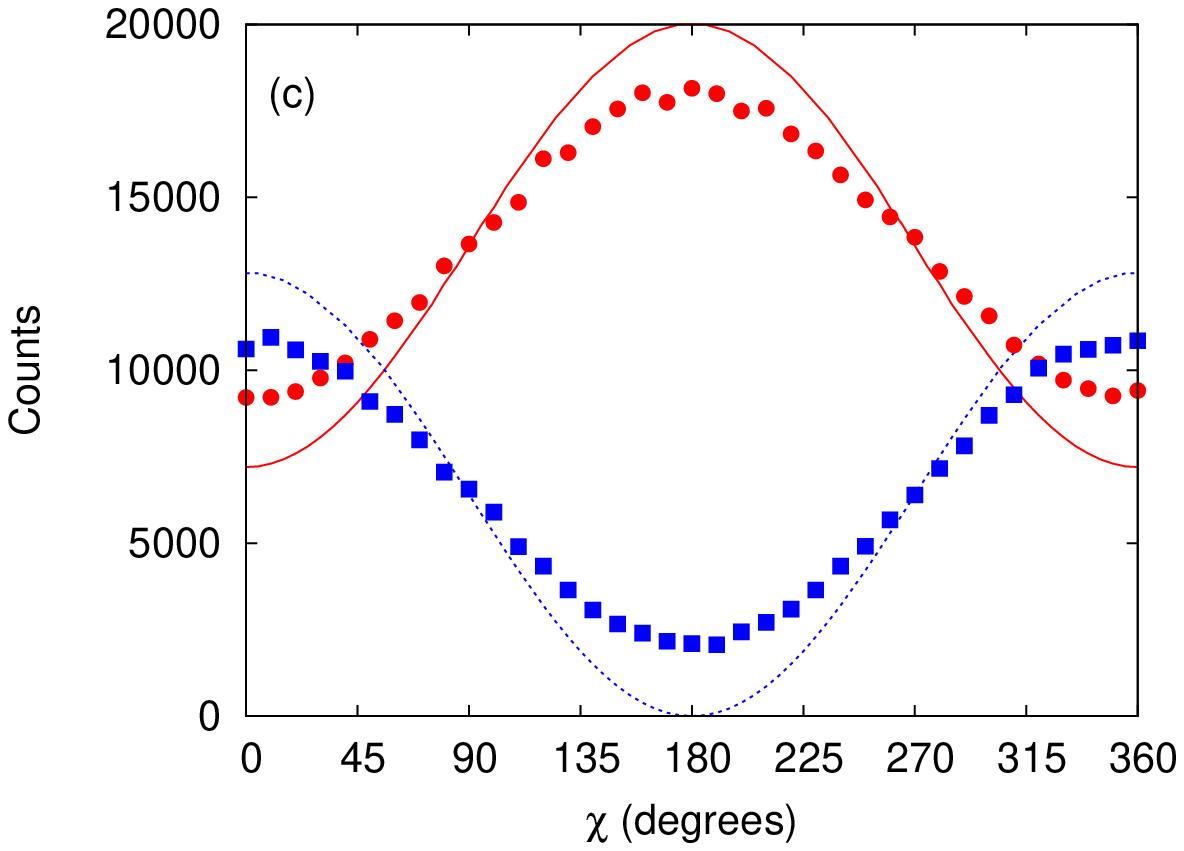}
\includegraphics[width=8cm ]{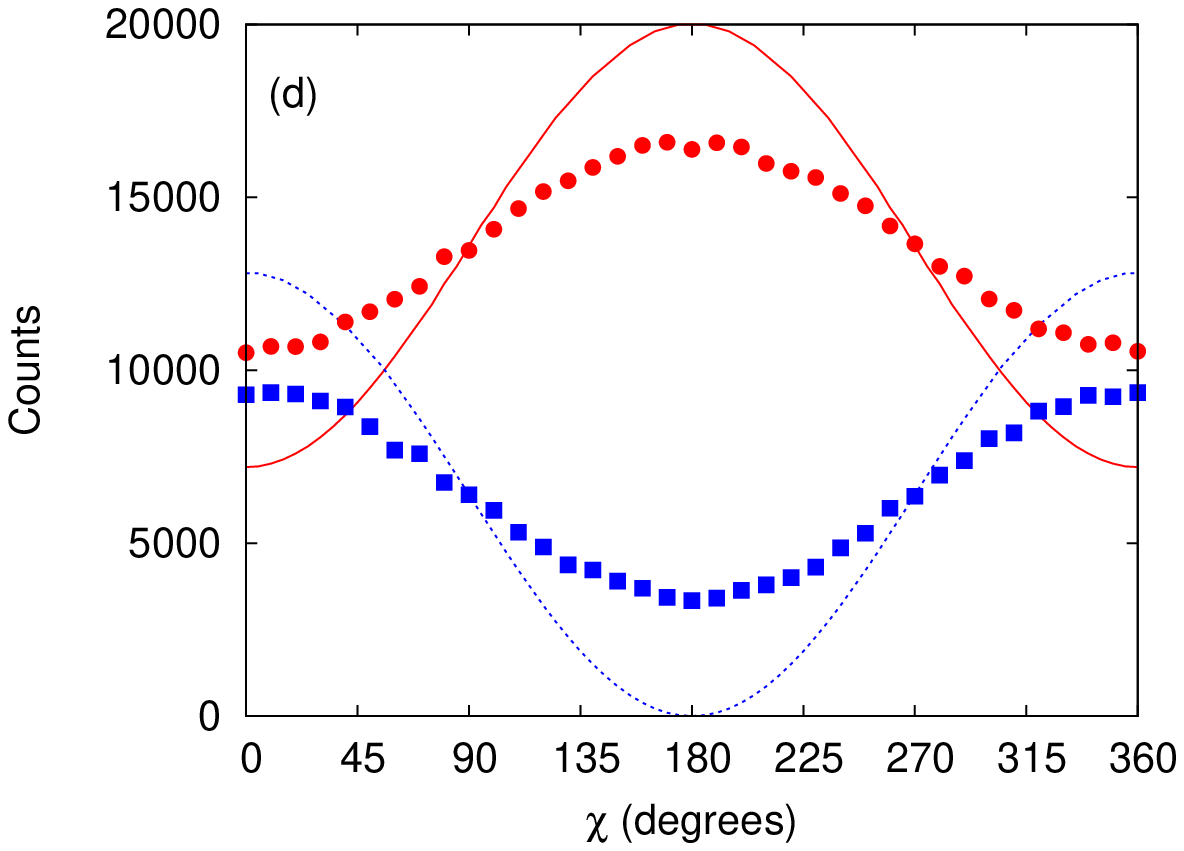}
\end{center}
\caption{%
Event-by-event simulation results of the
number of neutrons leaving the interferometer via the H-beam (red circles) and O-beam (blue squares)
as a function of the phase difference $\chi$ between the two paths inside the interferometer.
For each value of $\chi$, the number of particles generated in the simulation is $N=100000$.
The lines are the predictions of quantum theory for $a=b=1$.
Solid line: $p_\mathrm{H}$, see Eq.~(\ref{app2z});
dotted line: $p_\mathrm{O}$, see Eq.~(\ref{app2}).
(a)
Model parameters: reflection $R=0.2$, $\gamma=0.99$.
(b) Same as (a) except that $\gamma=0.5$,
reducing the accuracy and increasing the response time of the DLM.
(c)
Same as (a) except that to mimic the partial coherence of the incident neutron beam,
the initial message carried by each particle
has been modified by adding to $\psi^{(1)}$ and $\psi^{(2)}$
a random angle drawn uniformly from the interval $[-60^{\circ},60^{\circ}]$,
reducing the amplitude of the interference.
(d) Same as (c) except that $\gamma=0.5$.
}%
\label{figure.1}
\end{figure*}
\begin{figure*}[t]
\begin{center}
\includegraphics[width=12cm ]{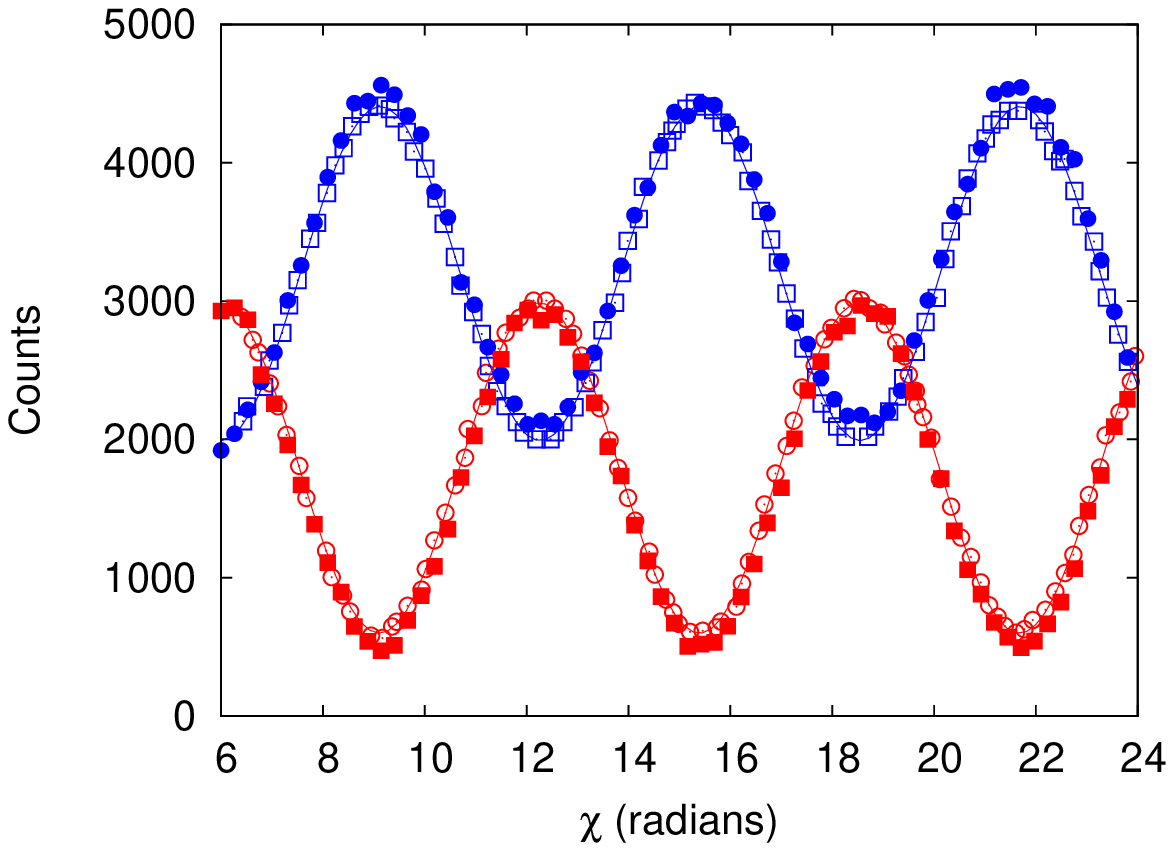}
\caption{%
Comparison between the data of a neutron interferometry experiment~\cite{KROU00} (open symbols)
and the results of an event-by-event simulation (solid symbols).
Open circles: counts per second and per square cm in the O-beam;
open squares: counts per second and per square cm in the H-beam;
solid circles: number of particles per sample leaving the interferometer via path 0;
solid squares: number of particles per sample leaving the interferometer via path 1.
The experimental data has been extracted from Fig.~2 of Ref.~\onlinecite{KROU00}.
The simulation parameters $R=0.22$ and $\gamma=0.5$ have been adjusted by hand to obtain a good fit
and the number of incident particles in the simulation is $N=22727$.
Lines through the data points are guides to the eye.
}%
\label{figure.9}
\end{center}
\end{figure*}

\subsection{Interferometer: Event-by-event simulation model}

Using the event-based processor described in Section~\ref{DLM},
it is straightforward to construct a simulation model for the interferometer shown in Fig.~\ref{qt-mzi}.
Without any modification, we use the event-based model of a beam splitter to simulate the
operation of BS0, BS1, BS2, and BS3.
Neutrons that are not refracted by BS1 or BS2 leave the apparatus and do not
contribute to the detection counts in the O- or H-beam.
During their flight from BS1 or BS2 to BS3, the neutrons pass through a metal foil which changes their time of flight~\cite{RAUC00}.
In the event-based model this effect of the metal foil is accounted for by
the phase shifters $\phi_0$ and $\phi_1$, see Fig.~\ref{qt-mzi}.
Thereby it is assumed that the absorption of neutrons by the metal foil is negligible~\cite{RAUC00}.
When the messenger passes through the phase shifter, its message changes according to
\begin{equation}
\mathbf{y}\leftarrow e^{i\phi_j}\mathbf{y}
,
\label{mess5}
\end{equation}
where $\phi_j$ represents the change in the time of flight as the neutron
passes through the metal foil on its way from BS1 to BS3 ($j=0$) or BS2 to BS3 ($j=1$).
In neutron interferometry experiments, minute rotations of the foils about an axis
perpendicular to the base plane of the interferometer induce large variations in $\phi_j$~\cite{RAUC00,LEMM10}.
All the neutrons which emerge from the interferometer through the O- or H-beam contribute
to the neutron count in these beams.

\subsection{Simulation results}

The simulation results presented in Fig.~\ref{figure.1}(a) demonstrate that the event-by-event simulation
reproduces the results of quantum theory if $\gamma$ approaches one~\cite{RAED05b,RAED05d,MICH11a}.
Indeed, there is excellent agreement with quantum theory.
In this example, the reflection coefficient of the beam splitters is taken to be $R=0.2$.
The parameter $\gamma$ which controls the learning pace of the DLM-based processor
can be used to account for imperfections of the neutron interferometer.
This is illustrated in Fig.~\ref{figure.1}(b) which shows simulation results for $\gamma=0.5$.

The quantum theoretical treatment of Section~\ref{qt1} assumes a fully coherent beam of neutrons.
In the event-based approach, the case of a coherent beam may be simulated by assuming
that the degree of freedom that accounts for the time of flight of the neutron
takes the same initial value each time a message is created.
In the event-based approach, we can mimic a partially coherent beam
by simply adding some random noise to the message,
that is when a message is created, a pseudo-random number in a specified range is added to $\psi^{(1)}$ and $\psi^{(2)}$.
In Fig.~\ref{figure.1}(c), we present simulation results for the case
that the random angle is drawn randomly and uniformly from the interval $[-\pi/3,\pi/3]$,
showing that reducing the coherence of the beam reduces the visibility,
as expected on the basis of wave theory~\cite{BORN64}.
Comparing Fig.~\ref{figure.1}(b) and Fig.~\ref{figure.1}(c), we conclude that
the same reduced visibility can be obtained by either
reducing $\gamma$ or by adding noise to the messages.
On the basis of this interferometry experiment alone, it is difficult to exclusively
attribute the cause of a reduced visibility to one of these mechanisms.
For completeness, Fig.~\ref{figure.1}(d) shows the combined effect
of decreasing $\gamma$ and adding noise to the messages on the visibility of the interference fringes.

Conclusive evidence that the event-based model reproduces the results of a
single-neutron interferometry experiment comes from comparing simulation data with experimental data.
In Fig.~\ref{figure.9}, we present such a comparison using
experimental data extracted from Fig.~2 of Ref.~\cite{KROU00}.
It was not necessary to try to make the best fit: the parameters $R$ and $\gamma$ and the offset
in $\chi$ were varied by hand.
As shown in Fig.~\ref{figure.9}, the event-based simulation model reproduces, quantitatively, the experimental results
reported in Fig.~2 of Ref.~\cite{KROU00}.

\section{Stochastic and deterministic beam attenuation}\label{blocking}
\begin{figure*}[t]
\begin{center}
\includegraphics[width=8cm ]{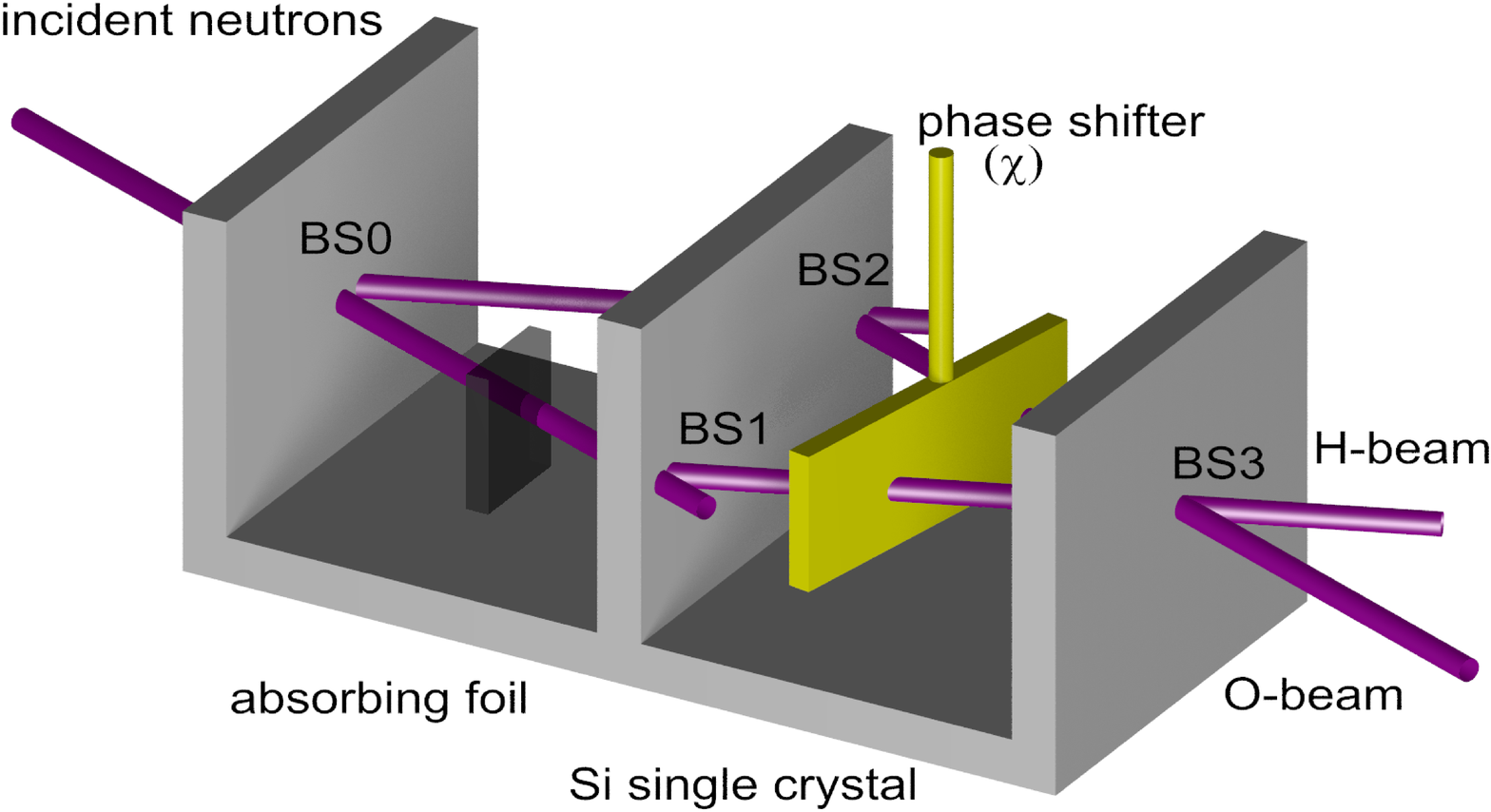}
\includegraphics[width=8cm ]{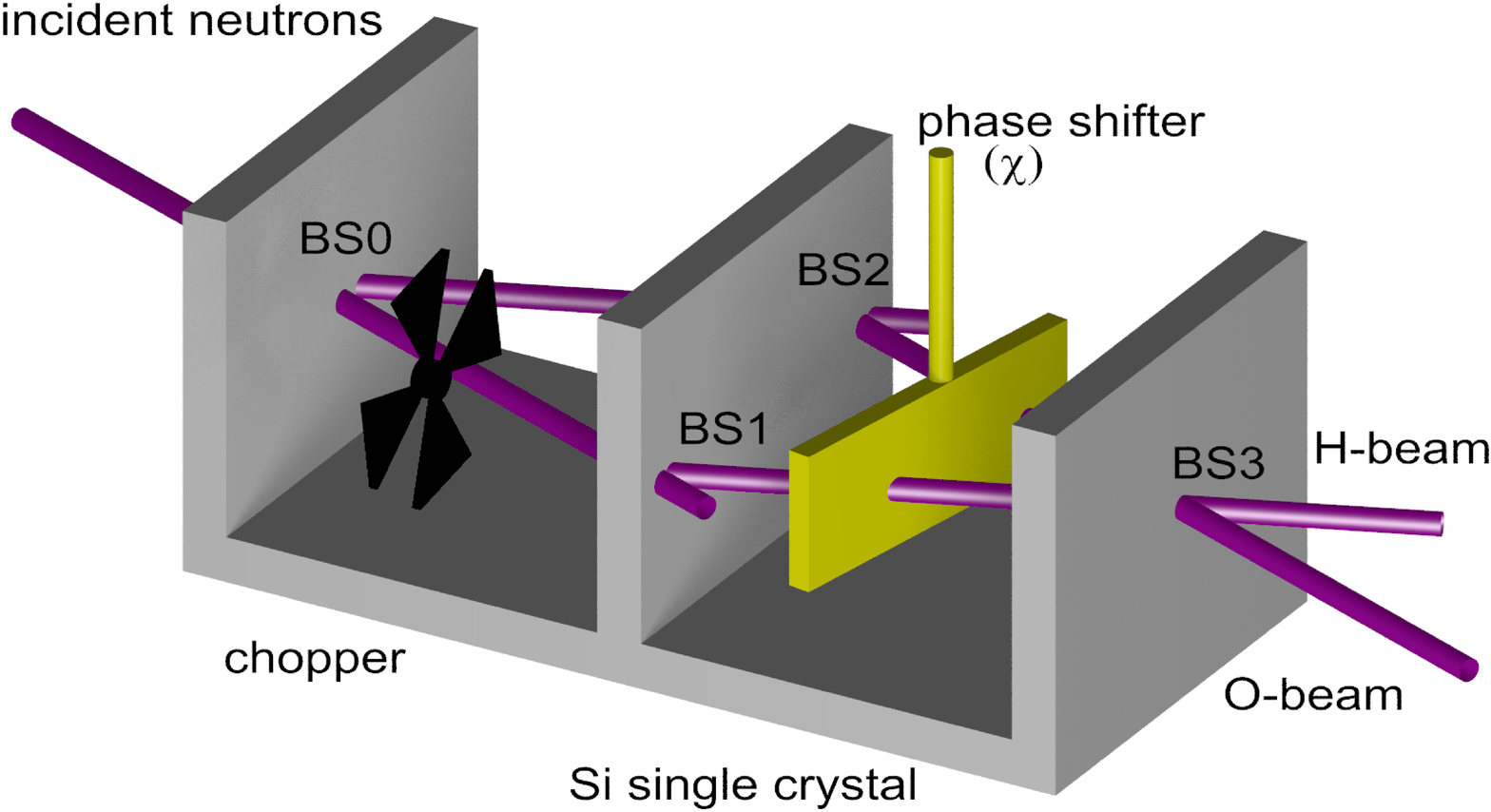}
\caption{%
Diagram of the single-neutron interferometry experiment
with a stochastic (left) and deterministic (right) absorber.
}%
\label{CHOPPERexperiment}
\end{center}
\end{figure*}

The second series of experiments that we consider are neutron interferometry experiments in which the beam from BS0 to BS1
is attenuated either by a partial, stochastic absorber~\cite{RAUC84a,RAUC00},
(see Fig.~\ref{CHOPPERexperiment}(a)),
an absorbing lattice~\cite{SUMM87,RAUC92,RAUC00}
or by a chopper, a rotating absorbing disc (see Fig.~\ref{CHOPPERexperiment}(b))
that periodically blocks neutrons from traveling to BS1~\cite{RAUC84a,SUMM87,RAUC92,RAUC00}.
We denote the average fraction of neutrons which pass the absorber/lattice/chopper by $0\le a\le 1$.

Let us assume that the incident flux of neutrons is constant in time.
Of all neutrons passing through the stochastic absorber, only the fractor $a$ ``survives''
the interaction with the absorber material.
Considering the case of the experiment with a chopper, as the rotation frequency of the chopper increases
up to the point that each individual neutron passes the chopper with probability $a$,
the difference between the deterministic and stochastic absorption is expected to disappear~\cite{RAUC84a,SUMM87,RAUC92,RAUC00,KALO92}.
Although in the experiment, the chopper rotates inside an aluminum chamber, the vibrations associated with
the rotation have an adverse effect on the amplitude of the interference fringes and hence,
the rotation frequency of the chopper was effectively limited to a few rotations per second~\cite{RAUC84a}.
This problem could partially be alleviated by using an absorbing lattice instead of a rotating disc~\cite{SUMM87,RAUC92,RAUC00}.
In this section, we assume that the rotation of the chopper does not affect the experimental outcomes
other than by stopping particles from reaching BS1.

\subsection{Quantum theory}

\begin{figure*}[t]
\begin{center}
\includegraphics[width=12cm ]{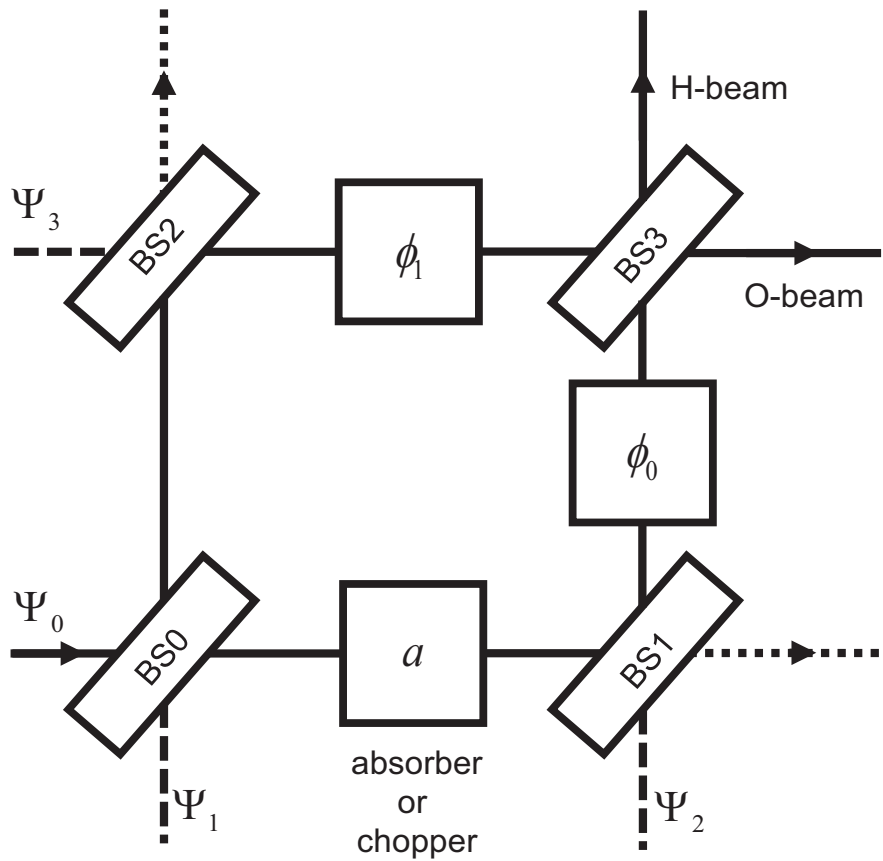}
\caption{%
Diagram of the single-neutron interferometry experiment with a stochastic or deterministic absorber~\cite{RAUC74a}.
BS0,...,BS3: beam splitters; $\phi_0$ and $\phi_1$: phase shifters.
Neutrons pass through the stochastic or deterministic absorber with probability $a$.
}%
\label{neutronChopper}
\end{center}
\end{figure*}
\begin{figure*}[t]
\begin{center}
\includegraphics[width=8cm ]{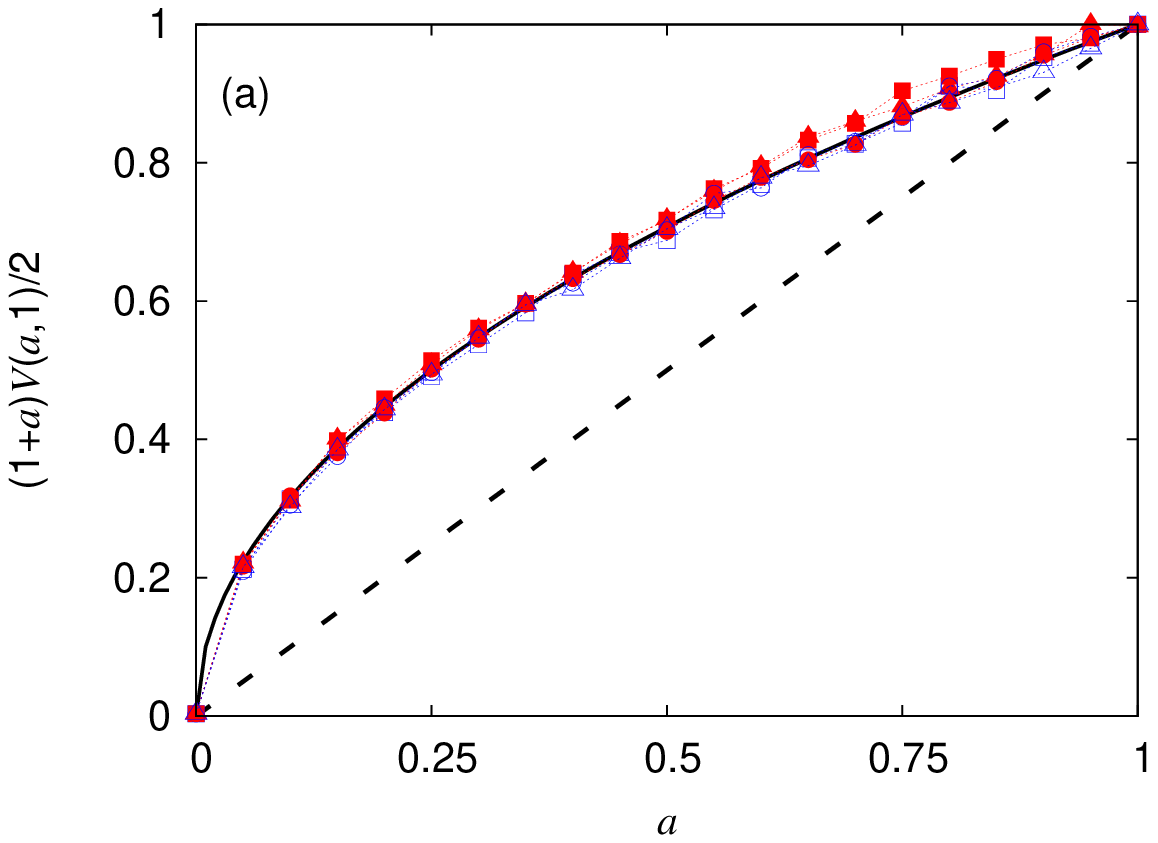}
\includegraphics[width=8cm ]{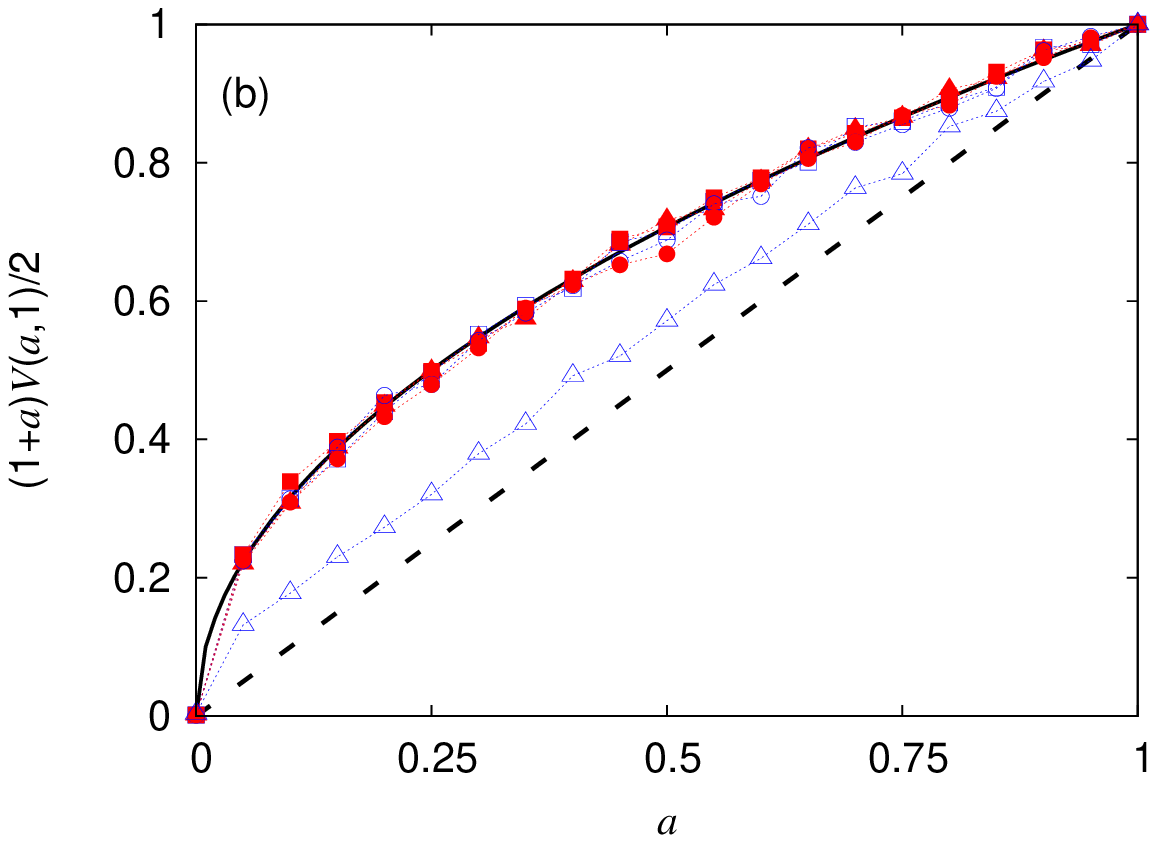}
\includegraphics[width=8cm ]{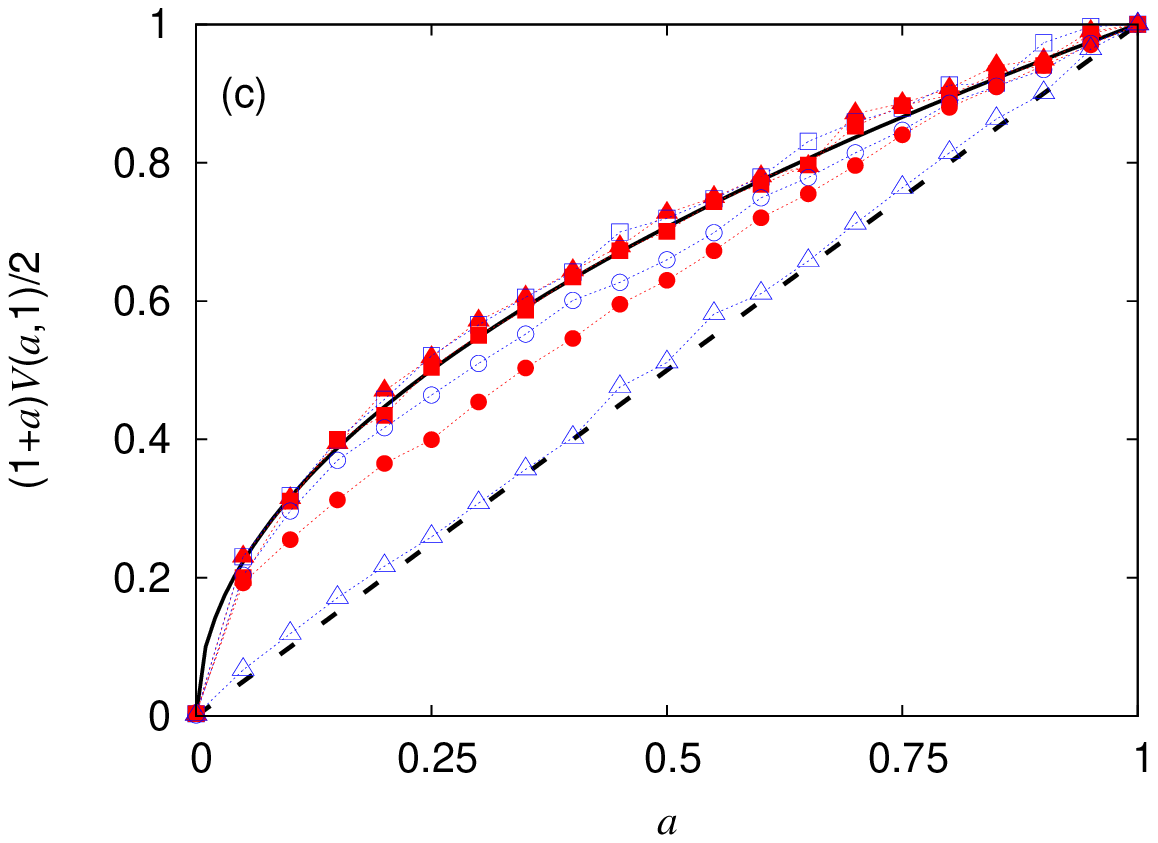}
\includegraphics[width=8cm ]{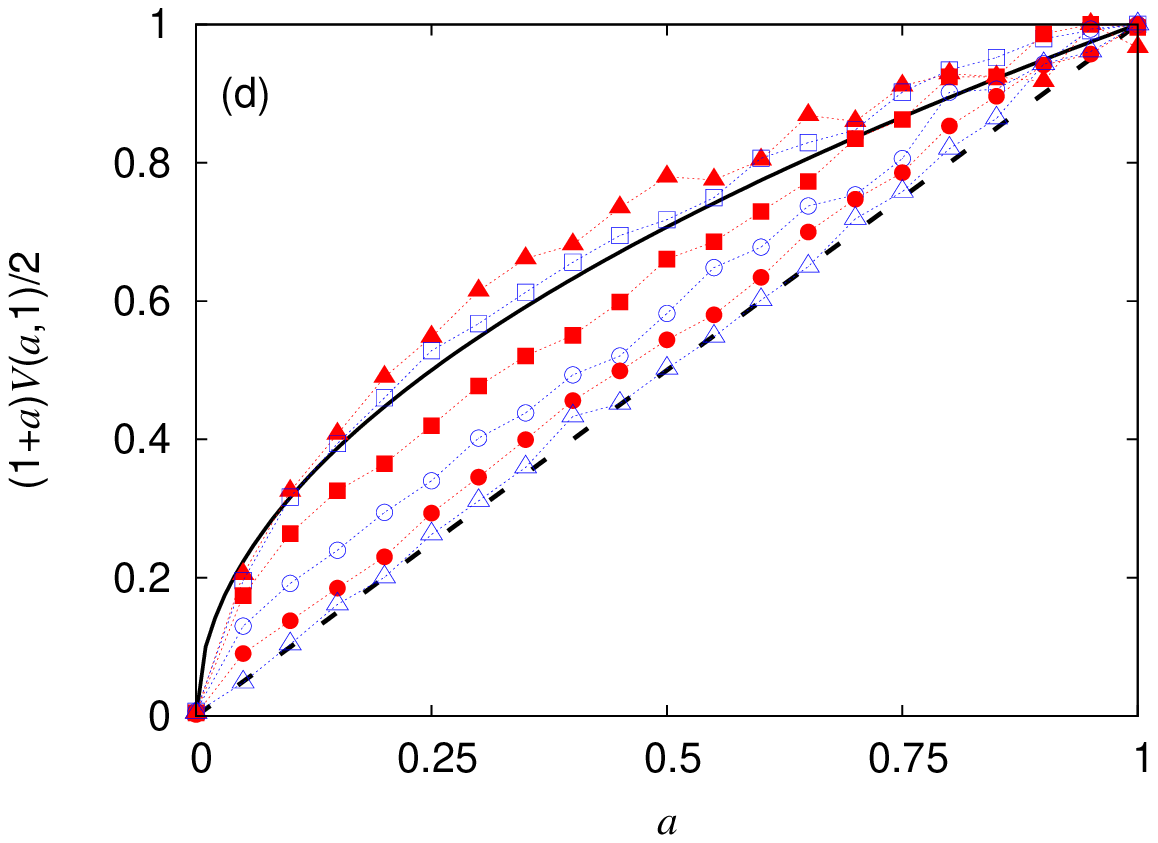}
\caption{%
Modulation amplitude of the intensity as a function of the probability $a$ that the neutrons pass the absorber,
obtained from event-by-event simulations of single-neutron interferometry experiments~\cite{RAUC84a,SUMM87,RAUC92}
with a stochastic (a) and a deterministic absorber (b,c,d)
for different values of $N_\mathrm{pc}$, the number of incident particles per cycle and per value of the phase shift $\chi$,
and $N_\mathrm{c}$, the number of cycles, keeping the total number $N_\mathrm{pc}N_\mathrm{c}$ fixed.
Solid line: square root dependence predicted by quantum theory for the case of a pure state (see text);
dashed line: linear dependence predicted by quantum theory for the case of a mixed state (see text);
solid triangles: $N_\mathrm{pc}=1000$, $N_\mathrm{c}=10$;
open squares: $N_\mathrm{pc}=100$, $N_\mathrm{c}=100$;
solid squares: $N_\mathrm{pc}=250$, $N_\mathrm{c}=40$;
open circles: $N_\mathrm{pc}=500$, $N_\mathrm{c}=20$;
solid circles: $N_\mathrm{pc}=1000$, $N_\mathrm{c}=10$;
open triangles: $N_\mathrm{pc}=10000$, $N_\mathrm{c}=1$.
Model parameters: reflection $R=0.2$,
$\gamma=0.98$ (a,b), $\gamma=0.9$ (c), $\gamma=0.5$ (d).
Within the statistical fluctuations, the simulation results for the stochastic absorber do not depend
on $N_\mathrm{pc}$ or $N_{c}$.
For all cases, increasing the number of cycles $N_\mathrm{c}$ reduces the statistical fluctuations only.
The maximum count of neutrons in the O-beam is less than 2000.
}%
\label{figure.5}
\end{center}
\end{figure*}

The case of the stochastic absorber has already been treated in Section~\ref{qt1}.
When the stochastic absorber is replaced by a chopper that either passes the neutrons
or blocks them completely, quantum theory prescribes that the experiment is described in terms of a mixed state~\cite{BALL03},
that is the observed intensity is the weighted sum of the two different experimental situations.
Denoting the average fraction of the neutrons which pass the chopper by $a$,
the probability and visibility of the O-beam are given by~\cite{RAUC84a}
\begin{eqnarray}
p'_\mathrm{O}&=&(1-a)p_\mathrm{O}(a=0,b=1)+ a p_\mathrm{O}(a=1,b=1)=TR^2\left(1+a+2a\cos\chi\right)
.
\label{app2b}
\end{eqnarray}
and
\begin{eqnarray}
{V}(a)&=&
\frac{\max_{\chi} p'_\mathrm{O} - \min_{\chi} p'_\mathrm{O}}{\max_{\chi} p'_\mathrm{O} + \min_{\chi} p'_\mathrm{O}}=\frac{2a}{1+a}
,
\label{app2c}
\end{eqnarray}
respectively.
In contrast to the case of the stochastic absorber where the modulation amplitude of the interference fringes is given by $\sqrt{a}$,
when a chopper is used to block neutrons that travel towards BS1,
the modulation amplitude of the interference fringes is given by $(1+a)V(a)/2=a$, that is it is linear in $a$~\cite{RAUC74a}.

In quantum theory, the difference between the stochastic and deterministic absorber enters through the choice of the state.
In the former case, we use a pure state to describe the interference pattern whereas
in the latter case, a mixed state which adds the probabilities for the two different experimental configurations
(chopper blocking or not).
However, this stationary-state wave theory does not include as a parameter the rotation speed of the chopper.
Only the fraction ($a$) of neutrons which pass with each cycle enters,
not how many neutrons pass through one opening in the disc at a time.
This problem calls for a solution of the time-dependent Schr\"odinger equation
but, to the authors knowledge, there is no report of successful work in this direction.
Put differently, in the quantum theoretical description of this experiment,
the parameter (chopper rotation frequency) which connects Eq.~(\ref{app2b}) to Eq.~(\ref{app2})
is lacking.

Regarding the individual neutron as a particle,
consider the case in which the chopper rotates very slowly relative to the pace with which the neutrons
arrive. Then, when the position of the chopper allows the neutrons to pass, many of them pass before the chopper closes
and during this period, we may expect to see the interference signal which is characteristic of the two-path interferometer.
If the chopper blocks the beam, there is no interference.
It is in this case that the mixed state describes the statistics of the experiment.

But what if the chopper rotates very fast such that with each open/close change, on average one neutron impinges on the chopper?
In this case, the detected signal is not simply the sum of two independent experiments (one with the
path from BS0 to BS1 blocked and another one with no blocked paths)
but should be the same as in the case of a stochastic absorber (with the corresponding value of $a$).
The next subsection shows that the event-based model effortlessly reproduces this behavior
and provides a unified, logically consistent description of these experiments.

\subsection{Event-based model}\label{mixedstate}
Both the case of a stochastic and time dependent absorber are easily incorporated in the event-based model
of the interferometer.
In the former case, particles leaving BS0 towards BS1 pass through the absorber if $\RN<a$
where $a$ is the fraction of particles that passes and $\RN$ is a uniform pseudo-random number
(which changes with each particle).
In the latter case, the procedure is as follows.
First, we define the (dimensionless) unit time interval by the time it takes for the chopper to open and close.
The number of neutrons incident on the interferometer per unit time interval
and the number of such intervals will be denoted by $N_{S}$ and $N_{I}$, respectively.
Thus, in one simulation run, the total number of neutrons created by the source is $N=N_{S}N_{I}$.
Uniform pseudo-random numbers are used to generate $N$ times in the interval $[0,N_I]$.
These are the times, relative to the motion of the chopper, at which the neutrons
will arrive at the chopper, if they followed the path from BS0 to BS1.
To a very good approximation, the interarrival times are distributed exponentially or,
in other words, the events are created according to a Poisson distribution~\cite{GRIM95}.
Neutrons are sent to the interferometer, one at a time, in chronological order.
When a neutron arrives at the chopper, its arrival time is used to determine if the
chopper is open or closed. If the chopper is open, the neutron continues its journey to BS1.
Otherwise, it is removed from the system and does not contribute to
the detection counts.
All neutrons that appear in the O- or H-beam are recorded by the detectors.

The data is collected as follows.
First we choose $0\le a \le 1$. Then, for each setting of the phase shift $\chi$,
the source sends $N=N_\mathrm{pc}N_\mathrm{c}$ neutrons to the interferometer.
Here, $N_\mathrm{pc}$ is the number of incident particles per cycle and per value of the phase shift $\chi$
and $N_\mathrm{c}$ is the number of cycles.
For a fair comparison between different cases, we keep $N$ fixed.
Increasing $N_\mathrm{c}$ while keeping $N_\mathrm{pc}$ constant reduces the statistical fluctuations only.
As in the case of the interferometer without absorbers,
the event-based processors in BS0, BS1, BS2 and BS3 perform their task, yielding counts in the O- and H-beam.
By changing $\chi$ and repeating the simulation, we determine the visibility $V(a)$.
Plotting $(1+a)V(a)/2$ as a function of $a$, we can directly
compare to the experimental results~\cite{RAUC84a,SUMM87,RAUC92,RAUC00}.
According to Eqs.~(\ref{app2a}) and (\ref{app2c}), $(1+a)V(a)/2$ is proportional
to $\sqrt{a}$ and $a$ for a pure and mixed state, respectively.

\subsection{Simulation results}

In the event-based approach, a certain number of particles has to pass through the interferometer before
an interference pattern appears (obviously, as in experiment, a single particle does not produce an interference pattern).
In addition, with each change of the chopper position, the learning machine
in BS3 has to adapt to the change in input
as particles arrive on either port 0 or 1 when the path via BS1 is not blocked
and no particles arrive on port 0 when the path via BS1 is blocked.
Thus, the learning machine should adapt quickly to a new situation,
a requirement which is in conflict with the desire to reproduce
the results of quantum theory, which demands $\gamma\rightarrow1^-$.
From these simple observations, we expect that the parameter $\gamma$
can be used, not only to control the visibility of the interference fringes (see Fig.~\ref{figure.1}(a,b))
but can also be used to produce features which in quantum theory, are characteristic of the mixed state.
The results presented in Fig.~\ref{figure.5} confirm these expectations.

In Fig.~\ref{figure.5}(a), we present the results for the stochastic absorber.
Disregarding the statistical fluctuations, the data nicely follow the $\sqrt{a}$ curve predicted by quantum theory
and is in agreement with experiment~\cite{RAUC84a,SUMM87,RAUC92,RAUC00}.
Note that in this case, as long as $N=N_\mathrm{pc}N_\mathrm{c}$ is fixed, the values of $N_\mathrm{pc}$ and $N_\mathrm{c}$ themselves
should, and also do not matter because the process to block neutrons from reaching BS1 is time independent.

Simulation results for the case of the deterministic absorber
are presented in Figs.~\ref{figure.5}(b,c,d), corresponding to $\gamma=0.98,0.9,0.5$, respectively.
For $\gamma=0.98$, see Fig.~\ref{figure.5}(b), the machines learn fairly slowly.
Yet for $N_\mathrm{pc}=10000$ (the value of $N_\mathrm{c}$ merely affects the statistical fluctuations),
the signature of the mixed state, the linear dependence of $(1+a)V(a)/2$ on $a$, starts to appear (open triangles).
For $\gamma=0.9$, see Fig.~\ref{figure.5}(c),
the simulation produces the results of both the pure and the mixed state (solid and open triangles, respectively).
Note that the event-based model also delivers intermediate results (solid circles),
as observed in experiment~\cite{SUMM87,RAUC92,RAUC00}.
For $\gamma=0.5$, see Fig.~\ref{figure.5}(d), the processors can quickly adapt to a new situation,
yielding results that interpolate smoothly between the linear- and square-law dependence on $a$.

Summarizing, the event-based model produces results that agree with the stochastic absorber or very
fast chopper, a fast chopper and a slow chopper, but unlike in the quantum theoretical descriptions,
without any modification to the simulation model but only
by changing the number of neutrons per open/close cycle, that is the rotation frequency of the chopper relative to the number
of incident neutrons.

\section{Violation of a Bell inequality}\label{violation}

\begin{figure*}[t]
\begin{center}
\includegraphics[width=16cm ]{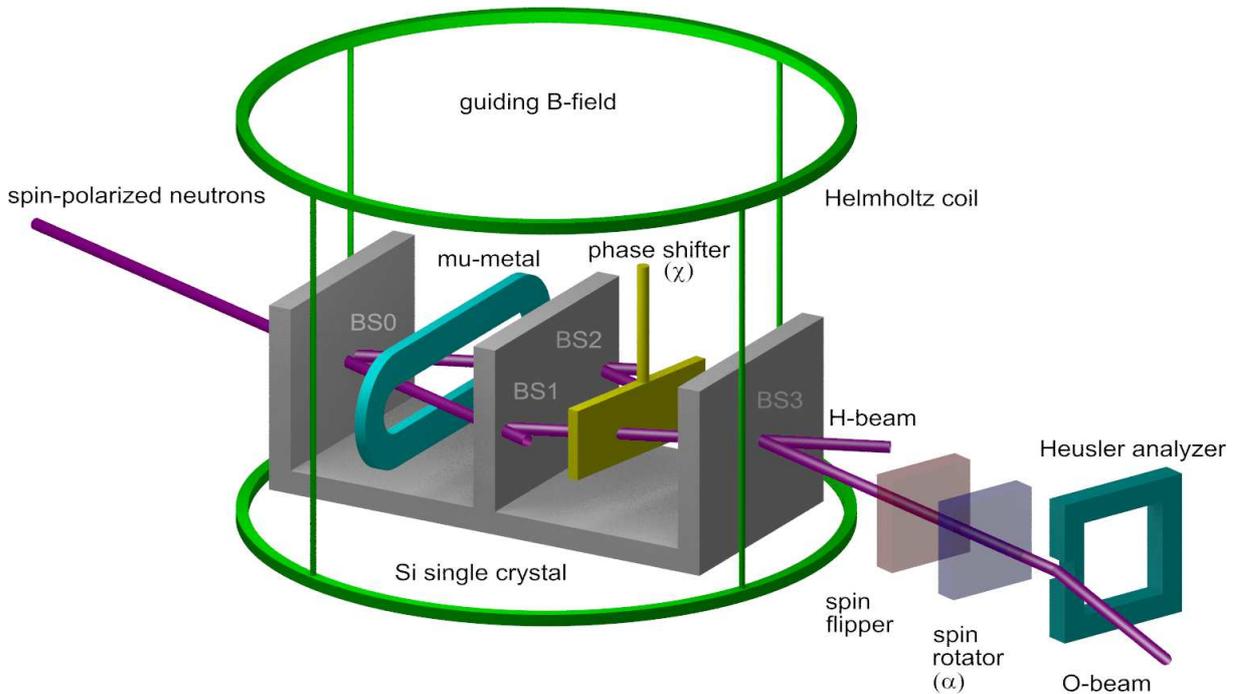}
\caption{%
Diagram of the single-neutron interferometry experiment to observe
correlations that cannot be accounted for by a quantum system in the product state,
see also Fig.~1 in Ref.~\cite{HASE03}.
BS0,...,BS3: beam splitters.
The combination of the mu-metal spin turner, phase shifter, and spin rotator allow
the independent manipulation of the neutron's spatial and magnetic degrees of freedom.
}%
\label{Bellexperiment}
\end{center}
\end{figure*}

The neutron interferometry experiment reported in Ref.~\cite{HASE03}
demonstrates that it is feasible to manipulate independently, the spatial
and spin degree of freedom of massive particles.
The experiment shows that it is possible to
create correlations between these two degrees of freedom which, within quantum theory,
cannot be described by a product state.
The direct experimental evidence is that the data for this correlation
violates a Bell-CHSH inequality~\cite{HASE03}.

In this section, we show that the event-based model faithfully reproduces all the features
of quantum theory for this experiment and, by changing the model parameter $\gamma$,
can also reproduce the numerical values of the correlations, as measured in experiments~\cite{HASE03,BART09}.

A diagram of the single-neutron interferometry experiment is shown in Fig.~\ref{Bellexperiment},
see also Fig.1 of Ref.~\cite{HASE03}.
Incident neutrons pass through a magnetic-prism polarizer (not shown)
that produces two spatially separated beams of neutrons with their magnetic moments aligned
parallel (spin up), respectively anti-parallel (spin down) with respect to the magnetic axis of the polarizer
which is parallel to the guiding field $\mathbf{B}$.
The spin-up neutrons impinge on a silicon-perfect-crystal interferometer.
On leaving beam splitter BS0, neutrons may or may not experience refraction.
A mu-metal spin-turner changes the orientation of the magnetic moment from parallel to perpendicular
to the guiding field $\mathbf{B}$. In detail, the result of passing through the
mu-metal spin-turner is that the magnetic moment of a neutron that travels towards BS1 (BS2)
rotates by $\pi/2$ ($-\pi/2$) about the $y$-axis.
Before the different paths join at the entrance plane of beam splitter BS3,
a difference between the times of flight (corresponding to a phase in the wave mechanical description)
along the two paths can be manipulated by a phase shifter.
The neutrons that experience two refraction events when passing through the interferometer
form the $\mathrm{O}$-beam and are analyzed by sending them through a spin rotator and
a Heusler spin analyzer. If necessary, to induce an extra spin rotation of $\pi$,
a spin flipper is placed between the interferometer and the spin rotator.
The neutrons that are selected by the Heusler spin analyzer are counted with a
neutron detector (not shown) that has a very high efficiency ($\approx 99\%$)~\cite{HASE03}.

\subsection{Quantum theory}

\begin{figure*}[t]
\begin{center}
\includegraphics[width=14cm ]{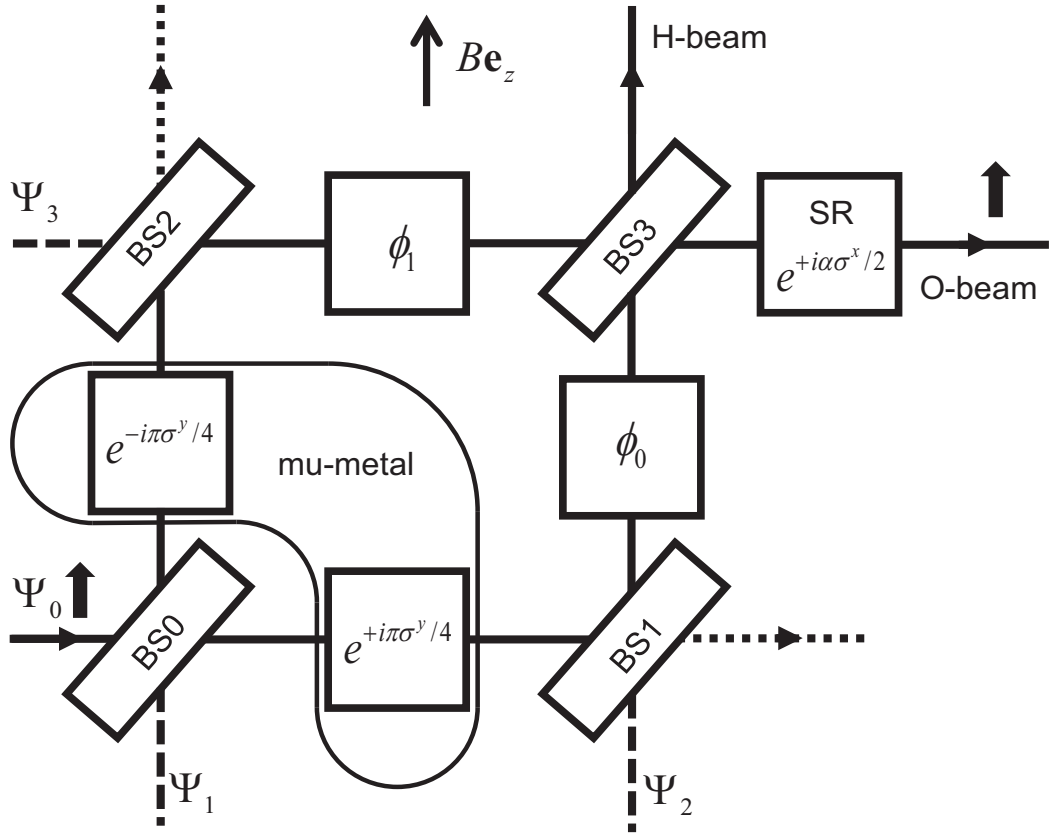}
\caption{%
Quantum theoretical model of the single-neutron Bell-inequality test experiment~\cite{HASE03}.
Polarizer spin-up neutrons are injected in the interferometer.
BS0,...,BS3: beam splitters;
Mu-metal: spin rotators in both paths;
$\phi_0$ and $\phi_1$: phase shifters;
SR: spin rotator by a variable angle $\alpha$;
$B\mathbf{e}_z$: constant magnetic field.
A detector counts all spin-up neutrons that leave the device via the O-beam.
Another detector counts all neutrons that leave the device via the H-beam.
}%
\label{qt-bell}
\end{center}
\end{figure*}

\begin{figure*}[t]
\begin{center}
\includegraphics[width=12cm ]{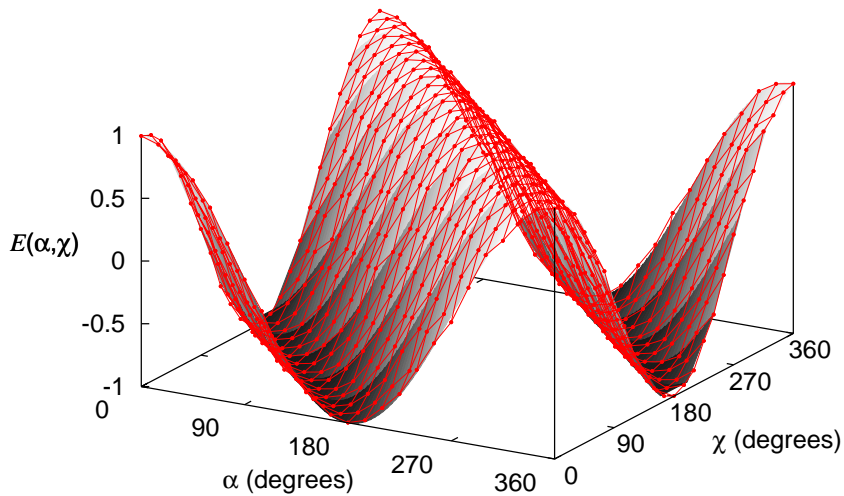}
\caption{%
Correlation $E(\alpha,\chi)$ between spin and path degree of freedom as obtained from
an event-by-event simulation of a single-neutron interferometry experiment which
shows violations of a Bell inequality~\cite{HASE03}.
Solid surface: $E(\alpha,\chi)=\cos(\alpha+\chi)$ predicted by quantum theory;
solid circles: simulation data. The lines connecting the markers are guides to the eye only.
Model parameters: reflection $R=0.2$ and $\gamma=0.99$.
For each pair $(\alpha,\chi)$, $N=10000$ particles were used to determine each of the four counts
$N(\alpha,\chi)$, $N(\alpha+\pi,\chi+\pi)$, $N(\alpha,\chi+\pi)$, and $N(\alpha+\pi,\chi+\pi)$
that appear in Eq.~(\ref{Exy}).
}%
\label{figure.2}
\end{center}
\end{figure*}
\begin{figure*}[t]
\begin{center}
\includegraphics[width=12cm ]{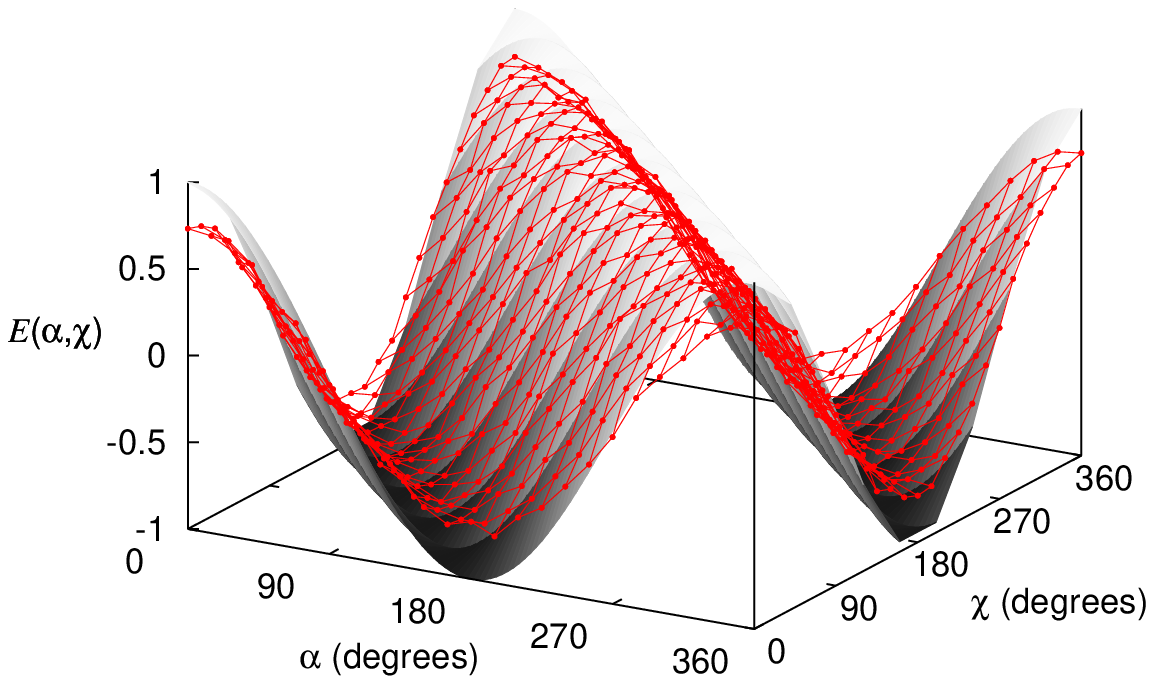}
\caption{%
Same as Fig.~\ref{figure.2} except that the
reflection $R=0.2$ and $\gamma=0.55$.
The differences between the quantum theoretical results and
the simulation data are due to the choice $\gamma=0.55$.
The event-based simulation reproduces the exact results of quantum theory if $\gamma\rightarrow1^-$ (data not shown).
}%
\label{figure.3}
\end{center}
\end{figure*}

An essential feature of the experiment is that as the neutron passes through the
interferometer, its path and its magnetic moment become correlated~\cite{HASE03}.
The quantum theoretical description of the experiment reported in Ref.~\cite{HASE03}
requires a four-state system for the path and another two-state system to account for the spin-1/2 degree-of-freedom.
Thus, the statistics of the experimental data is described by the state vector Eq.~(\ref{app0}).

In the experiment~\cite{HASE03}, the neutrons that enter the interferometer
have their spins up, relative to the direction of the guiding field $\mathbf{B}$ (see Fig.~\ref{Bellexperiment}).
Thus, the state describing the incident neutrons is $|\Psi\rangle=(1,0,0,0,0,0,0,0)^T$, omitting irrelevant
phase factors.
As the state vector propagates through the interferometer and the spin rotator (see Fig.~\ref{qt-bell}),
it changes according to
\begin{eqnarray}
|\Psi'\rangle
&=&
\left(\begin{array}{cc}
        \phantom{i}\cos(\alpha/2)& i\sin(\alpha/2) \\
        i\sin(\alpha/2) & \phantom{i}\cos(\alpha/2)
\end{array}\right)_{6,7}
\left(\begin{array}{cc}
        \phantom{-}\T^\ast &\R\\
        -\R^\ast & \T
\end{array}\right)_{5,7}
\left(\begin{array}{cc}
        \phantom{-}\T^\ast &\R \\
        -\R^\ast & \T
\end{array}\right)_{4,6}
\left(\begin{array}{cc}
         e^{i\phi_1}&0 \\
         0 & e^{i\phi_1}
\end{array}\right)_{6,7}
\left(\begin{array}{cc}
         e^{i\phi_0}&0 \\
         0 & e^{i\phi_0}
\end{array}\right)_{4,5}
\nonumber \\ &&\times
\left(\begin{array}{cc}
        \phantom{-}\T^\ast &\R\\
        -\R^\ast & \T
\end{array}\right)_{3,7}
\left(\begin{array}{cc}
        \phantom{-}\T^\ast &\R \\
        -\R^\ast & \T
\end{array}\right)_{2,6}
\left(\begin{array}{cc}
        \T & -\R^\ast \\
        \R & \phantom{-}\T^\ast
\end{array}\right)_{1,5}
\left(\begin{array}{cc}
        \T & -\R^\ast \\
        \R & \phantom{-}\T^\ast
\end{array}\right)_{0,4}
\nonumber \\ &&\times
\left(\begin{array}{cc}
        \phantom{-}1/\sqrt{2} & \phantom{-}1/\sqrt{2} \\
        -1/\sqrt{2} & \phantom{-}1/\sqrt{2} \\
\end{array}\right)_{2,3}
\left(\begin{array}{cc}
        \phantom{-}1/\sqrt{2} & -1/\sqrt{2} \\
        \phantom{-}1/\sqrt{2} & \phantom{-}1/\sqrt{2}
\end{array}\right)_{0,1}
\left(\begin{array}{cc}
        \T & -\R^\ast \\
        \R & \phantom{-}\T^\ast
\end{array}\right)_{1,3}
\left(\begin{array}{cc}
        \T & -\R^\ast \\
        \R & \phantom{-}\T^\ast
\end{array}\right)_{0,2}
|\Psi\rangle
,
\label{app5}
\end{eqnarray}
where the subscripts $i,j$ refer to the pair of elements of the
eight-dimensional vector on which the matrix acts.
Reading backwards, the first pair of matrices in Eq.~(\ref{app5}) represents beam splitter BS0,
the second pair the mu-metal (a spin rotation about the $y$-axis by $\pi/4$ and $-\pi/4$, respectively),
the third and fourth pair beam splitters BS1 and BS2, respectively,
the fifth pair the phase shifters,
the sixth pair beam splitter BS3,
and the last matrix represents the spin rotator SR.

From Eq.~(\ref{app5}), it follows that the probability to detect a neutron with spin up in the O-beam is given by
\begin{eqnarray}
p_\mathrm{O}(\alpha,\chi)&=&|\Psi'_{3,\uparrow}|^2=TR^2  \left[1+\cos(\alpha+\chi)\right]
,
\label{app6}
\end{eqnarray}
where $\chi=\phi_0-\phi_1$.
From Eq.~(\ref{app6}) it follows that the correlation $E_\mathrm{O}(\alpha,\chi)$ is given by~\cite{HASE03}
\begin{eqnarray}
E_\mathrm{O}(\alpha,\chi)&\equiv&\frac{%
p_\mathrm{O}(\alpha,\chi)+p_\mathrm{O}(\alpha+\pi,\chi+\pi)-p_\mathrm{O}(\alpha+\pi,\chi)-p_\mathrm{O}(\alpha,\chi+\pi)
}{
p_\mathrm{O}(\alpha,\chi)+p_\mathrm{O}(\alpha+\pi,\chi+\pi)+p_\mathrm{O}(\alpha+\pi,\chi)+p_\mathrm{O}(\alpha,\chi+\pi)
}
=\cos(\alpha+\chi)
,
\label{app7}
\end{eqnarray}
independent of the reflection $R=|r|^2=1-T$ of the beam splitters (which have been assumed to be identical).
Repeating the calculation for the probability of detecting a neutron in the H-beam shows
that $E_\mathrm{H}(\alpha,\chi)=0$, independent of the direction of the spin.
Note that if the mu-metal would rotate the spin about the $x$-axis instead of about the $y$-axis,
we would find $E_\mathrm{O}(\alpha,\chi)=\cos\alpha\cos\chi$, a typical expression for a quantum system in a product state.

The fact that $E_\mathrm{O}(\alpha,\chi)=\cos(\alpha+\chi)$ implies that the state of the neutron
cannot be written as a product of the state of the spin and the phase.
In other words, in quantum parlance, the spin- and phase-degree-of-freedom are entangled~\cite{BASU01,HASE03}.
In this context, it is customary to form the Bell-CHSH function~\cite{BELL93,CLAU69}
\begin{eqnarray}
S=S(\alpha_1,\chi_1,\alpha_2,\chi_2)&=&
 E_\mathrm{O}(\alpha_1,\chi_1)
+E_\mathrm{O}(\alpha_1,\chi_2)
-E_\mathrm{O}(\alpha_2,\chi_1)
+E_\mathrm{O}(\alpha_2,\chi_2)
,
\label{app8}
\end{eqnarray}
for some set of experimental settings $\alpha_1$, $\chi_1$, $\alpha_2$, and $\chi_2$.
If the quantum system can be described by a product state, we must have $|S|\le2$.
Therefore, if experiment shows that $|S|>2$, it is impossible
to interpret the experimental result in terms of a quantum system in the product state~\cite{BALL03}.
If $\alpha_1=0$, $\chi_1=\pi/4$, $\alpha_2=\pi/2$, and $\chi_2=\pi/4$, then
$S=2\sqrt{2}$, the maximum value allowed by quantum theory~\cite{CIRE80}.

The single-neutron interferometry experiment yields
the count rate $N(\alpha,\chi)$ for the spin-rotation angle $\alpha$ and the difference $\chi$
of the phase shifts of the two different paths in the interferometer~\cite{HASE03}.
Following Ref.~\cite{HASE03}, the correlation $E(\alpha,\chi)$
is defined by
\begin{eqnarray}
\label{Exy}
E(\alpha,\chi)&=&
\frac{N(\alpha,\chi)+N(\alpha+\pi,\chi+\pi)-N(\alpha,\chi+\pi)-N(\alpha+\pi,\chi)}{N(\alpha,\chi)+N(\alpha+\pi,\chi+\pi)+N(\alpha,\chi+\pi)+N(\alpha+\pi,\chi)}
,
\end{eqnarray}
and, if quantum theory describes this experiment, we expect that $E(\alpha,\chi)\approx E_\mathrm{O}(\alpha,\chi)$.
Experiments show that $S>2$~\cite{HASE03,BART09}.

\subsection{Event-by-event model: realization}\label{BELLrealization}

The components that constitute the interferometer have been described in Section~\ref{interferometer}.
In the following, we specify the action of the remaining components,
namely the magnetic-prism polarizer (not shown), the mu-metal spin-turner,
the spin-rotator and spin analyzer.

\medskip\noindent
{\bf Magnetic-prism polarizer:}
this device takes as input a neutron with an unknown magnetic moment
and produces a neutron with a magnetic moment that is either parallel (spin up) or anti-parallel (spin down)
with respect to the $z$-axis (which by definition is parallel to the guiding field $\mathbf{B}$).
In the experiment, only a neutron with spin up is injected into the interferometer.
Therefore, to simplify matters a little, we assume that the neutrons
that enter the interferometer all have spin up.
This assumption is easily incorporated in the procedure that creates the initial message by
simply setting $\theta=0$ (see Eq.~(\ref{mess2})).

\medskip\noindent
{\bf Mu-metal spin-turner:}
the action of this component is to rotate the magnetic moment of the neutron
by $\pi/2$ ($-\pi/2$) about the $y$-axis, depending on whether the neutron was transmitted
(reflected) by BS0.
The processor that accomplishes this is very simple.
It takes as input the direction of the magnetic moment, represented by the message $\mathbf{y}$
and performs the rotation
$\mathbf{y}\leftarrow e^{i\pi\sigma^y/4}\mathbf{y}$
which corresponds to a rotation about the $y$-axis by $\pi/2$.
We emphasize that we use Pauli matrices as a convenient tool to express rotations in 3D space,
not because in quantum theory the magnetic moment of the neutron is represented by spin-1/2 operators.

\medskip\noindent
{\bf Spin-rotator:}
the action of this component is to rotate the magnetic moment of a neutron
by an angle $\alpha$ about the $x$-axis.
It changes the message according to
$\mathbf{y}\leftarrow e^{i\alpha\sigma^x/2}\mathbf{y}$.

\medskip\noindent
{\bf Spin analyzer:}
this component selects neutrons with spin up, after which these neutrons are counted by a 100\%
efficient detector.
The simplest algorithm that performs this task is to project the magnetic moment
on the $z$-axis and send the neutron to the detector if the projected value
exceeds a pseudo-random number $\RN$.

\medskip\noindent
{\bf Detector:} We simply count all neutrons that appear in the O- and H-beam.

\subsection{Simulation results}

In Fig.~\ref{figure.2} we present results for the correlation Eq.~(\ref{Exy}) as obtained
from event-by-event simulations of the experimental setup depicted in Fig.~\ref{qt-bell},
assuming that the experimental conditions are very close to ideal.
For the ideal experiment, quantum theory predicts that $E(\alpha,\chi)=\cos(\alpha+\chi)$
(represented by the solid surface in Fig.~\ref{figure.2})
and as shown by the markers in Fig.~\ref{figure.2}, disregarding the small statistical fluctuations,
there is close-to-prefect agreement between the event-by-event simulation data and quantum theory.

The real experiment suffers from unavoidable imperfections, leading to a reduction
and distortion of the interference patterns~\cite{HASE03}.
In an event-by-event approach, it is easy to incorporate mechanisms for
different sources of imperfections by modifying or adding rules.
After all, we can manipulate each individual event.
However, to reproduce the available data, this is not necessary because,
as before, we can use the parameter $\gamma$ to control the deviation from the quantum theoretical result.
In particular, we can use the parameter $\gamma$ to fit the simulation results to the experimental data for the value of $S$.

For instance, taking $R=0.2$ and $\gamma=0.55$, the simulation (see Fig.~\ref{figure.3}) yields  $S_\mathrm{max}=2.05$,
in excellent agreement with the value $2.052\pm0.019$ obtained in experiment~\cite{HASE03}.
For $R=0.2$ and $\gamma=0.67$, the simulation yields  $S_{max}=2.30$,
in very good agreement with the value $2.291\pm0.008$ obtained in a similar,
more recent experiment~\cite{BART09}.

\subsection{Discussion}\label{discussion}

An Einstein-Podolsky-Rosen-Bohm (EPRB) experiment
can be used to test for violations of a Boole-Bell-type inequality
but not all experiments that test for violations of a Boole-Bell-type inequality
are EPRB experiments.
Essential features of an EPRB (thought) experiment are that
(1) a source emits pairs of particles with properties of which at least one is correlated,
(2) as the particles leave the source, they no longer interact (but the correlations of their properties do not change),
(3) the properties of the particles are determined by two spatially separated analyzers which do not communicate with each other
and have settings that may be changed independently and randomly,
(4) all particles leaving the source are analyzed and contribute to the averages and correlations.
Clearly, the neutron interferometry experiment which we have discussed in this section is not an EPRB experiment.
It only satisfies the fourth criterion if we disregard the neutrons that
are transmitted by BS1 or BS2.

The violation of the Bell-CHSH inequality observed in the neutron interferometry experiment
demonstrates that it is possible to create a correlation between the path and the magnetic
moment of the neutron, although there is no direct ``interaction'' between the two.
If we interpret the outcome of this experiment in terms of quantum theory,
the observed violation of the Bell-CHSH inequality implies that it is impossible
to describe the outcome of the experiment in terms of a product state
of path and spin states. Hence, the system must be described by an entangled state.
On the other hand, a classical, Einstein-local and causal, event-by-event
process can also reproduce all the features of the entangled state.
Hence, not too much significance should be attached to the latter.

For completeness, we mention that if we pick the angle $\chi$
randomly from the same finite set of predetermined values used
to produce Fig.~\ref{figure.2},
an event-based simulation with $\gamma=0.99$ yields (within the usual
statistical fluctuations) the correlation
$E(\alpha,\chi)\approx (1/2)\cos(\alpha+\chi)$, which
does not lead to a violation of a Bell-type inequality (data not shown).
Thus, if the neutron interferometry experiment could be
repeated with random choices for the position
of the phase shifter ($\chi$),
and the experimental results would show a significant violation of
a Bell-type inequality, the event-based model that we have
presented here would be ruled out.

\begin{figure*}[t]
\begin{center}
\includegraphics[width=14cm ]{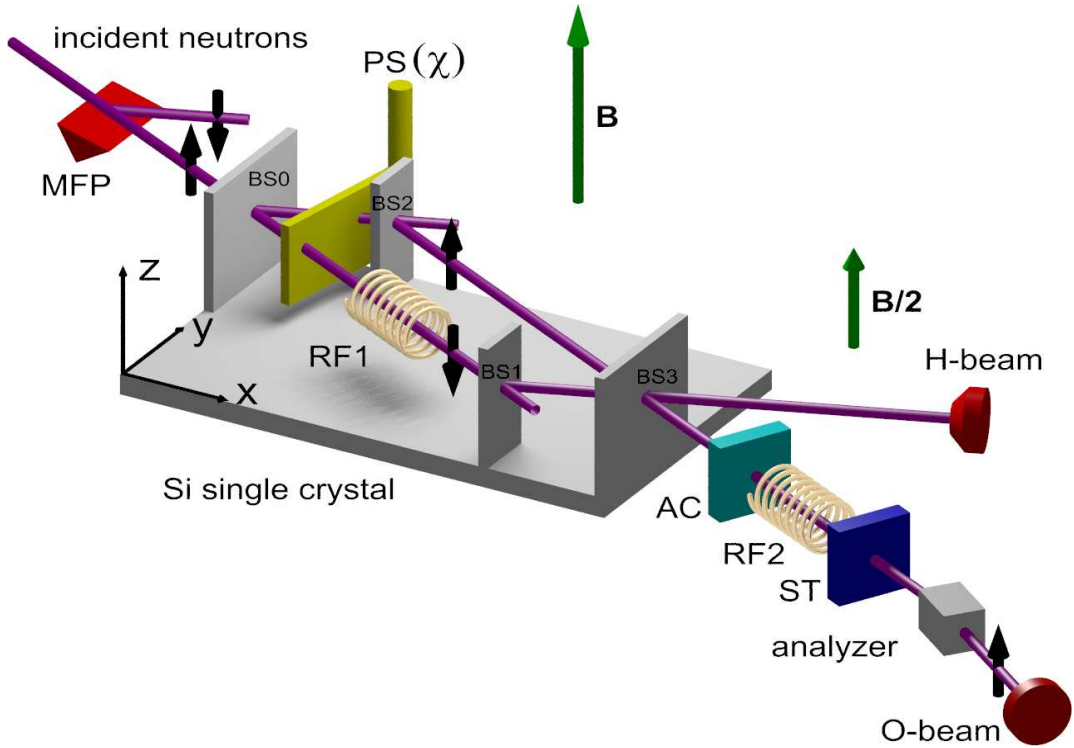}
\caption{%
A single-neutron interferometry experiment for the observation of interference between
the phases induced by a phase shifter (PS) and by two RF fields~\cite{SPON08}.
BS0,...,BS3: beam splitters;
MFP: magnetic field prism;
$\mathbf{B}=B\mathbf{e}_z$: constant magnetic field;
PS: phase shifter ;
RF1, RF2: radio-frequency spin-flippers;
AC: accelerator coils;
ST: spin turner.
}%
\label{SPONARexperiment}
\end{center}
\end{figure*}

\begin{figure*}[t]
\begin{center}
\includegraphics[width=14cm ]{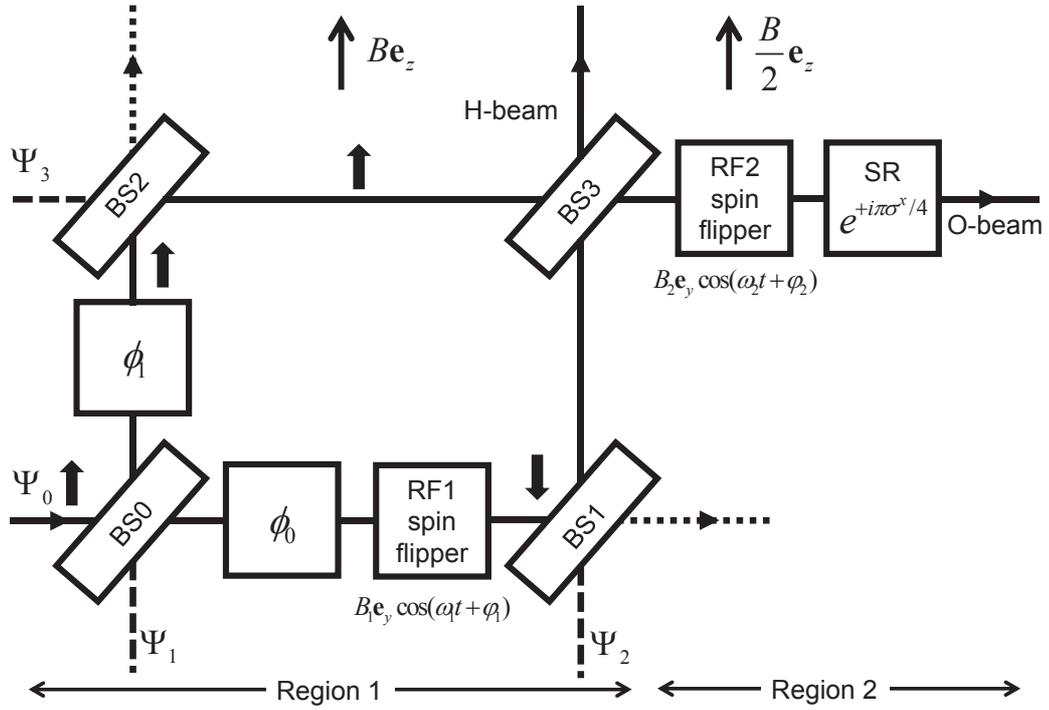}
\caption{%
Quantum theoretical model of the single-neutron interferometry experiment that demonstrates
the entanglement of the spin, path and energy degrees of freedom~\cite{SPON08}.
Polarizer spin-up neutrons are injected in an interferometer.
BS0,...,BS3: beam splitters;
$\phi_0$ and $\phi_1$: phase shifters;
RF1, RF2 spin flippers: radio frequency coils, tuned to the resonance frequencies
of the neutron spin in the static magnetic fields $B$ and $B/2$, respectively;
SR: spin rotator (magnet causing a magnetic moment to rotate about the $x$-axis by $\pi/2$).
A detector counts the spin-up neutrons in the O-beam.
Another detector counts all neutrons in the H-beam.
}%
\label{qt-RF}
\end{center}
\end{figure*}

\begin{figure*}[t]
\begin{center}
\includegraphics[width=12cm ]{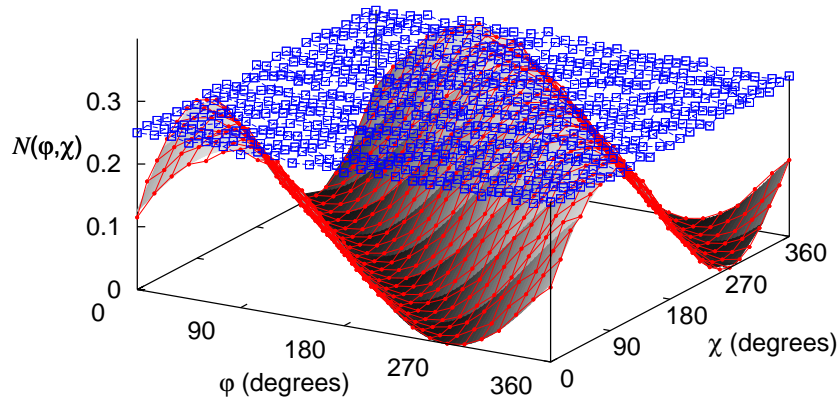}
\caption{%
Normalized counts $N(\varphi,\chi)$ as obtained from
an event-by-event simulation of a single-neutron interferometry experiment that
employs two radio-frequency fields to manipulate the energy of a single neutron~\cite{SPON08}.
Solid surface: probability $p_\mathrm{O}(\varphi,\chi)=(1-R)R^2[1+\sin(\chi+\varphi)]$
to observe neutrons in the O-beam, as predicted by quantum theory;
solid circles and open squares: simulation data for the normalized neutron counts in the O- and H-beam, respectively.
The lines connecting the markers are guides to the eye only.
Quantum theory predicts the probability to observe a neutron in the H-beam to be $R(R^2+(1-R)^2)=1/4$.
Model parameters: reflection $R=0.5$, $\gamma=0.99$;
number of particles per data point $N=10000$.
}%
\label{figure.6}
\end{center}
\end{figure*}
\begin{figure*}[t]
\begin{center}
\includegraphics[width=12cm ]{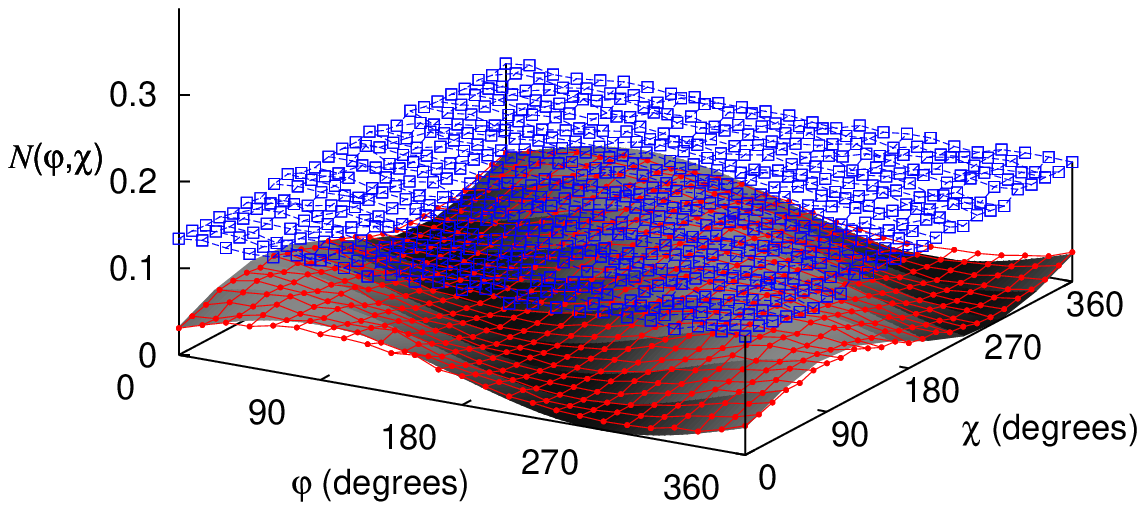}
\caption{%
Same as Fig.~\ref{figure.6} except that $R=0.2$ and $\gamma=0.5$.
Quantum theory predicts the probability to observe a neutron in the H-beam to be $R(R^2+(1-R)^2)\approx0.19$.
Differences between the quantum theoretical results and
the simulation data are due to the choice $\gamma=0.5$.
The event-based simulation reproduces the exact results of quantum theory if $\gamma\rightarrow1^-$ (data not shown).
}%
\label{figure.7}
\end{center}
\end{figure*}

\section{Coherent manipulation of the neutron spin, phase and energy}\label{manipulation}

In the language of quantum theory, the neutron interferometry experiment described in Ref.~\cite{SPON08}
demonstrates that using magnetic resonance techniques, it is possible to create entanglement between the
spin, path and energy degree of freedom of the neutron.

The set-up of the single-neutron interferometry experiment for the observation of
interference between the phases induced by a phase shifter (PS) and by two radio-frequency (RF) fields
is shown in Fig.~\ref{SPONARexperiment} (see also Fig.~1 in Ref.~\cite{SPON08}).
Neutrons with their spin polarized along the static magnetic field $\mathbf{B}=B\mathbf{e}_z$
impinge on the interferometer.
The radio-frequency coil (RF1) within one path of the skew-symmetric
neutron interferometer is tuned such that it flips the spin from up to down
and induces a phase shift ($\varphi_1$).
Neutrons that leave the interferometer in the H-beam are counted by a neutron detector.
Neutrons that leave the interferometer in the O-beam pass an
accelerator coil (AC) that compensates for the differences in times of flight of the two
different paths in the interferometer.
These neutrons pass through
a second radio-frequency coil (RF2) which is placed in a region where
the static magnetic field is $\mathbf{B}/2$ and which is driven by the half frequency of that of RF1.
RF2 flips the spin and induces a phase shift ($\varphi_2$).
Subsequently, neutrons with their spin up are counted by the combination
of a $\pi/2$ spin rotator (SR), an analyzer and a neutron detector.

\subsection{Quantum theory}

In Fig.~\ref{qt-RF}, we show the diagram that corresponds to
the experiment~\cite{SPON08}.
Spin-polarized neutrons impinge on beam splitter BS0.
A constant magnetic field ($B\mathbf{e}_z$) defines the spin-up direction.
The Hamiltonian that describes the spin of the neutron
as it moves through the radio-frequency coil RF1 reads
\begin{eqnarray}
H_1&=&-B \mathbf{e}_z S^z- B_1\mathbf{e}_y S^y \cos(\omega t + \varphi_1)
,
\label{app10}
\end{eqnarray}
where $B$ and $B_1$ are the static and radio-frequency (RF) fields
and $\omega$ and $\varphi_1$ denote the angular frequency
and phase of the RF field, respectively.
The wave function of the neutron spin, denoted by $|\Psi(t)\rangle$
satisfies the time-dependent Schr\"odinger equation (TDSE)
\begin{eqnarray}
i\frac{\partial}{\partial t} |\Psi(t)\rangle&=&H_1(t)|\Psi(t)\rangle
,
\label{app11}
\end{eqnarray}
where from now on, we use units such that $\hbar=1$.
Writing $|\Psi(t)\rangle=e^{i(\omega t+\varphi_1)S^z} e^{iB_1tS^y/2}|\Phi(t)\rangle$,
and imposing the resonance condition $B=\omega$,
the TDSE for $|\Phi(t)\rangle$ reads
$i\partial|\Phi(t)\rangle/\partial t =H^\prime_1(t)|\Phi(t)\rangle$ where
\begin{eqnarray}
H^\prime_1(t)&=&
-\frac{B_1}{4}\left[
S^y\cos2(\omega t+\varphi_1)-S^x \sin2(\omega t+\varphi_1)\cos\frac{B_1t}{2}
- S^z \sin2(\omega t+\varphi_1)\sin\frac{B_1t}{2}
\right]
,
\label{app12}
\end{eqnarray}
showing that $H^\prime_1(t)$ oscillates with frequencies $2\omega$ and $2\omega\pm B_1/2$.
We assume that the effect of these oscillations on the spin can be neglected, as is usually assumed
in NMR/ESR theory~\cite{ABRA61,ERNS90,WEIL94}.
As the neutron spin passes through the spin flipper RF1, its time evolution is given by the unitary matrix
\begin{eqnarray}
U_1(t)&=&e^{i(\omega t_1+\varphi_1)S^z} e^{iB_1t_1S^y/2}
,
\label{app13}
\end{eqnarray}
where $t_1$ is the time during which the neutron experiences the RF field of RF1.
Adjusting $B_1$ such that $B_1t_1/2=\pi$, the RF field
changes the neutron spin from up to down (see Fig.~\ref{qt-RF})
and changes the phase by $\omega t_1+\varphi_1$.

As the neutron leaves BS3, its spin can be up or down.
All neutrons that leave BS3 via the H-beam are sent to a detector.
Neutrons with spin up (down) that are transmitted (reflected) by BS3
fly through the spin-flipper RF2, operating at $\omega_2=\omega/2$
at resonance with the static field $B/2$ that is present in region 2.
The two RF-spin-flippers and the static
fields in regions 1 and 2 act as an interferometer
for the energy of the neutron~\cite{SPON08}.
Finally, the neutrons pass a spin rotator SR which
rotates the magnetic moment of the neutron about the $x$-axis by $\pi/2$,
mixing the spin-up and spin-down components.

As the state vector propagates through the interferometer and the spin analyzer, see
Fig.~\ref{qt-RF}, it changes according to
\begin{eqnarray}
|\Psi'\rangle
&=&
\left(\begin{array}{cc}
        \phantom{i}1/\sqrt{2}& i/\sqrt{2} \\
        i/\sqrt{2} & \phantom{i}1/\sqrt{2}
\end{array}\right)_{6,7}
\left(\begin{array}{cc}
        0 &\phantom{-}e^{i(\omega t_2/2+\varphi_2)/2} \\
        {-}e^{i(\omega t_2/2+\varphi_2)/2} & 0
\end{array}\right)_{7,8}
\left(\begin{array}{cc}
        \phantom{-}\T^\ast &\R\\
        -\R^\ast & \T
\end{array}\right)_{5,7}
\left(\begin{array}{cc}
        \phantom{-}\T^\ast &\R \\
        -\R^\ast & \T
\end{array}\right)_{4,6}
\nonumber \\ &&\times
\left(\begin{array}{cc}
        \phantom{-}\T^\ast &\R\\
        -\R^\ast & \T
\end{array}\right)_{3,7}
\left(\begin{array}{cc}
        \phantom{-}\T^\ast &\R \\
        -\R^\ast & \T
\end{array}\right)_{2,6}
\left(\begin{array}{cc}
        \T & -\R^\ast \\
        \R & \phantom{-}\T^\ast
\end{array}\right)_{1,5}
\left(\begin{array}{cc}
        \T & -\R^\ast \\
        \R & \phantom{-}\T^\ast
\end{array}\right)_{0,4}
\left(\begin{array}{cc}
        1 &0 \\
        0 &e^{i\omega   t_\downarrow}
\end{array}\right)_{4,5}
\nonumber \\ &&\times
\left(\begin{array}{cc}
        0 &\phantom{-}e^{i(\omega t_1+\varphi_1)/2} \\
        {-}e^{i(\omega t_1+\varphi_1)/2} & 0
\end{array}\right)_{0,1}
\left(\begin{array}{cc}
         e^{i\phi_1}&0 \\
         0 & e^{i\phi_1}
\end{array}\right)_{2,3}
\left(\begin{array}{cc}
         e^{i\phi_0}&0 \\
         0 & e^{i\phi_0}
\end{array}\right)_{0,1}
\left(\begin{array}{cc}
        \T & -\R^\ast \\
        \R & \phantom{-}\T^\ast
\end{array}\right)_{1,3}
\left(\begin{array}{cc}
        \T & -\R^\ast \\
        \R & \phantom{-}\T^\ast
\end{array}\right)_{0,2}
|\Psi\rangle
.
\label{app14}
\end{eqnarray}
Reading backwards, the matrices in Eq.~(\ref{app14}) represent
beam splitter BS0 (first pair of matrices),
the phase shifters inducing a relative phase shift $\chi=\phi_0-\phi_1$ (second pair) ,
the RF-spin-flipper RF1 (fifth matrix),
a phase shift (sixth matrix) applied to the spin-down neutron
($  t_\downarrow$ is the time during which a neutron appearing in the O-beam has spin down)
beam splitters BS1, BS2 and BS3 (matrices 7 through 12),
the RF-spin-flipper RF2 (matrix 13),
and finally the spin rotator SR (a spin rotation about the $x$-axis by $\pi/2$).
The analyzer (not included in Eq.~(\ref{app14})) sends the spin-up neutrons in the O-beam to the detector.

From Eq.~(\ref{app14}), it follows that the probability to detect a neutron with spin up in the O-beam
is given by
\begin{eqnarray}
p_\mathrm{O}(\varphi,\chi )&=&|\Psi'_{3,\uparrow}|^2= T R^2
\left\{1+\sin\left[\chi+\omega  t_\downarrow+(t_2-t_1)\omega/2+\varphi\right]\right\}
,
\label{app15}
\end{eqnarray}
where $\varphi=\varphi_2-\varphi_1/2$,
showing that the spin ($\varphi_1$,$\varphi_2$), path ($\chi$),
and energy ($\omega  t_\downarrow$) degree-of-freedom
are entangled and can be manipulated independently.
In the experiment~\cite{SPON08}, the phases induced by the guiding fields in regions 1 and 2 and
the phase $\omega  t_\downarrow+(t_2-t_1)\omega/2$ were compensated for by a tunable
accelerator coil with a static magnetic field along the $z$-direction~\cite{SPON08}.
Putting $\omega  t_\downarrow+(t_2-t_1)\omega/2=0$ in Eq.~(\ref{app15}) yields
\begin{eqnarray}
p_\mathrm{O}(\varphi,\chi )&=&|\Psi'_{3,\uparrow}|^2=
TR^2\left\{1+\sin\left[\chi+\varphi\right]\right\}
.
\label{app16}
\end{eqnarray}
The probability to find a neutron leaving the device via the H-beam reads
\begin{eqnarray}
p_\mathrm{H}&=&|\Psi'_{2,\uparrow}|^2+|\Psi'_{2,\downarrow}|^2= R(T^2+R^2)
,
\label{app17}
\end{eqnarray}
which obviously does not depend on $\chi$, $\varphi_1$, $t_1$, $\varphi_2$, or $t_2$.

\subsection{Event-by-event model: realization}

The components that constitute the interferometer have been described in Section~\ref{interferometer}
and the components that manipulate the neutron spin (in the rotating frame in the case of the RF flippers)
have been described in Section~\ref{BELLrealization}.
Therefore, we simply use these components without modification.
Also the simulation procedure is the same as before.
The source sends a particle
to BS0 and it is not until  that particle has left the interferometer or has been detected
in either the O- or H-beam that the source will send the next particle.

\subsection{Simulation results}

In Fig.~\ref{figure.6},
we present results for the normalized particle count $N(\varphi,\chi )$ in the
O-beam (solid circles) and H-beam (open squares) in the case of the ideal experiment.
The normalization consists of dividing the actual count by the number of particles $N$
generated by the source.
The quantum theoretical expression for the normalized O-beam count is given
by Eq.~(\ref{app16}) and is represented by the solid surface in Fig.~\ref{figure.6}.
According to Eq.~(\ref{app17}), quantum theory predicts that the normalized H-beam count
is independent of $\chi$ and $\varphi$.
From Fig.~\ref{figure.6}, it is obvious that the event-based simulations
reproduces all the features of the quantum theoretical description of this experiment.

The effect of changing the reflection from $R=0.5$ (Fig.~\ref{figure.6}) to $R=0.2$
in combination with reducing the parameter $\gamma$ is shown in Fig.~\ref{figure.7}.
As expected from Eqs.~(\ref{app16}) and (\ref{app17}), the normalized counts are reduced
and, because $\gamma=0.5$ instead of being close to one, the simulation data deviate (slightly)
from the quantum theoretical results (solid surface) Eq.~(\ref{app16}).

Our conclusion is that the event-based particle-only model reproduces the results of quantum theory for
a neutron interferometry experiment which, in the language of quantum theory, exhibits entanglement~\cite{SPON08}.

\begin{figure*}[t]
\begin{center}
\includegraphics[width=7cm ]{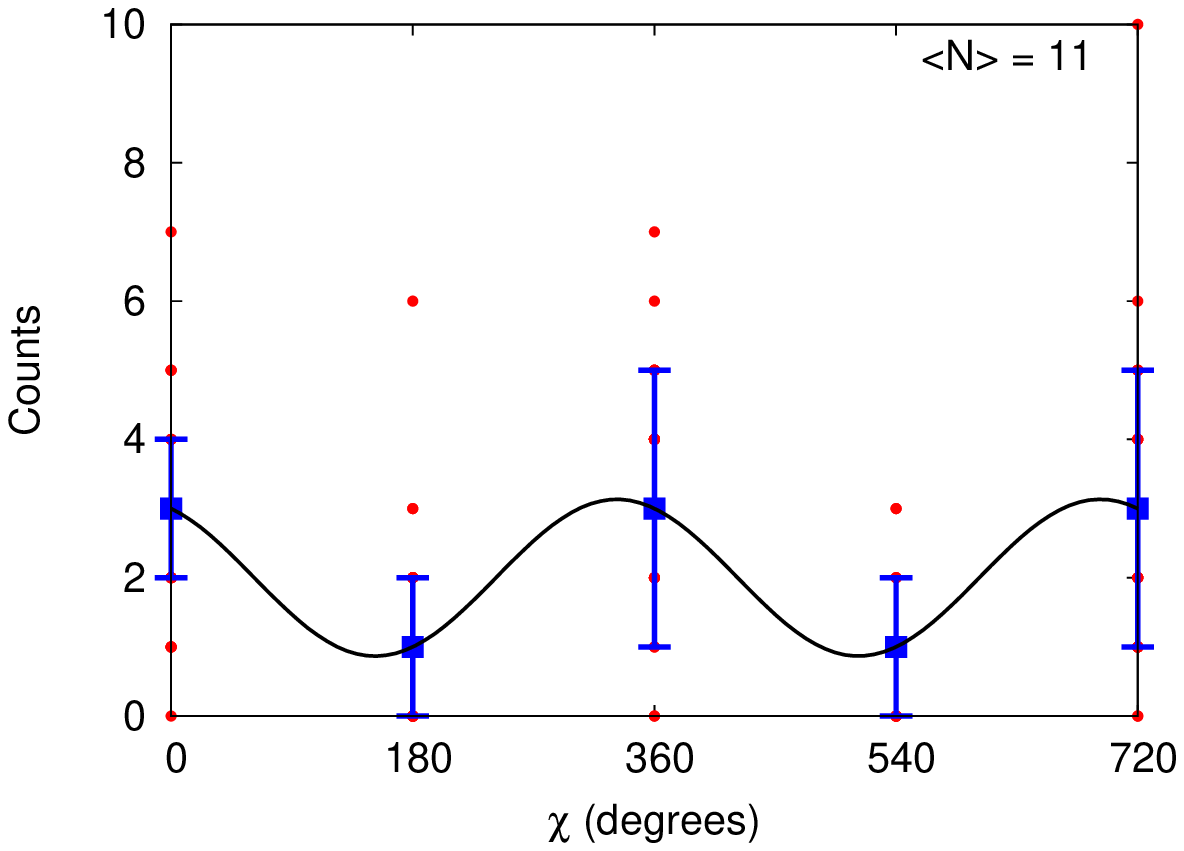}
\includegraphics[width=7cm ]{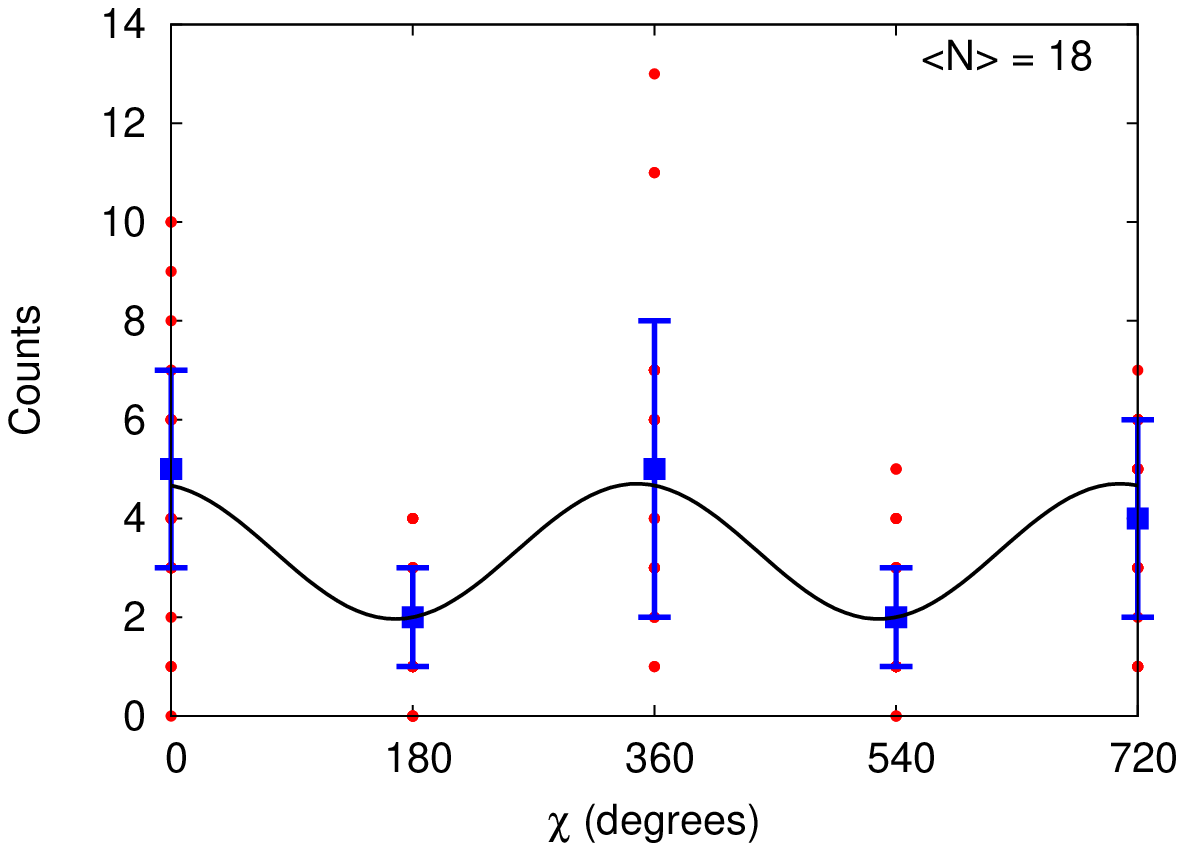}
\includegraphics[width=7cm ]{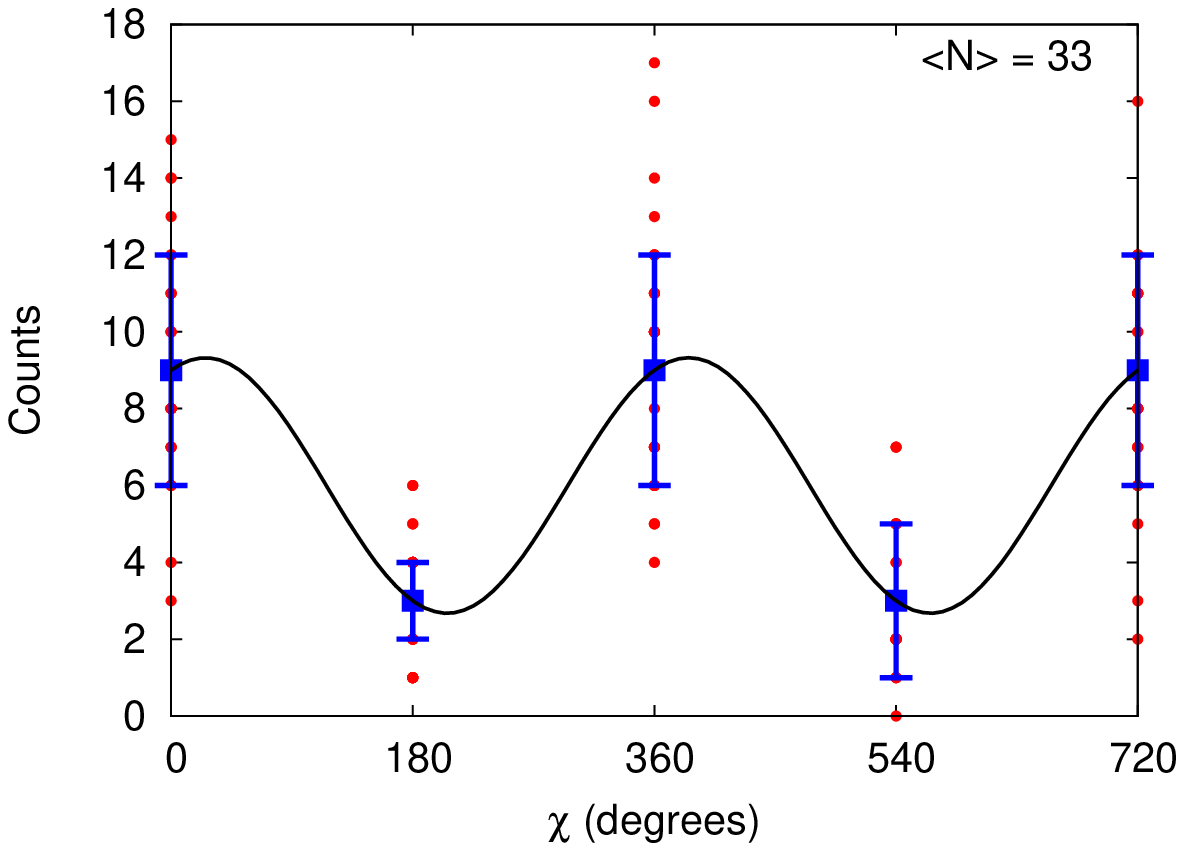}
\includegraphics[width=7cm ]{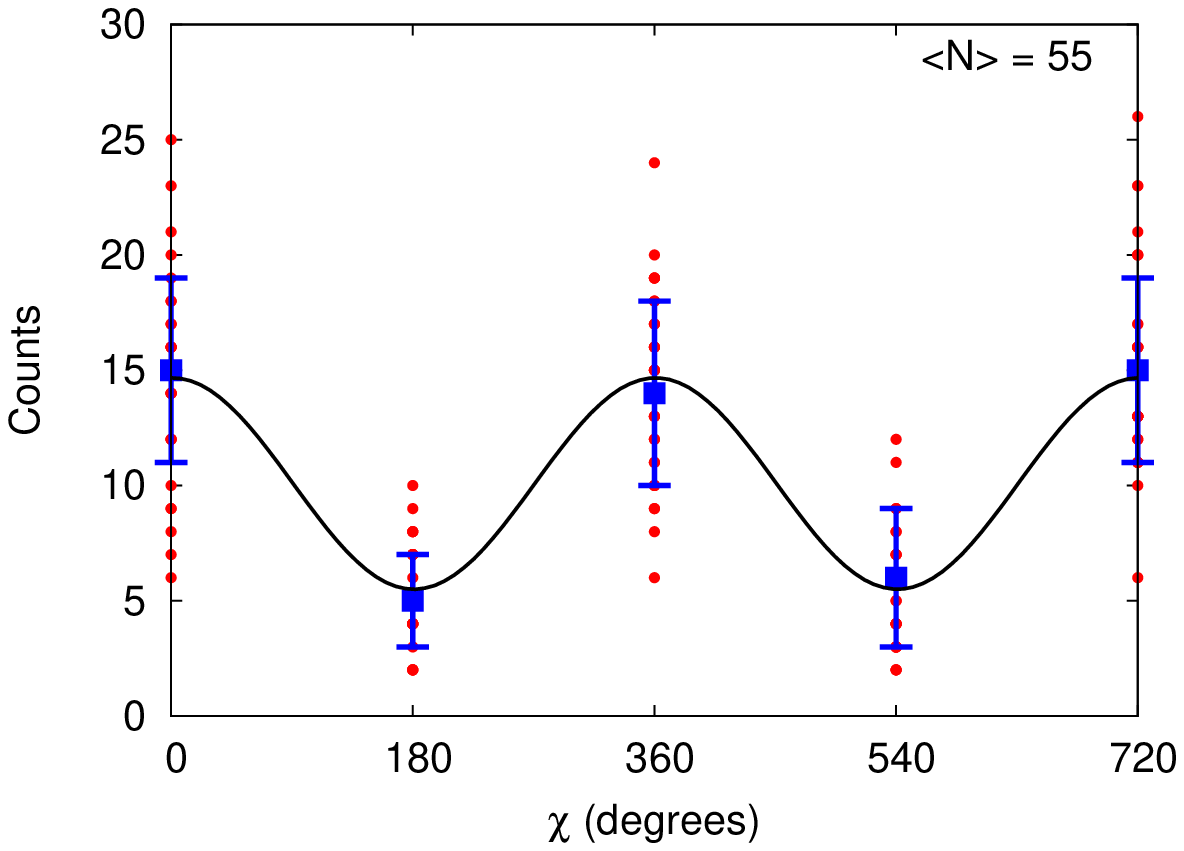}
\caption{%
Phase shift dependence extracted from interference patterns taken with a small number of neutrons,
see also Fig.~5 in Ref.~\onlinecite{RAUC90}.
The number of the detected neutrons in the O-beam, summed over the five different settings
of the phase shift and averaged over 30 independent simulations
is $\langle N\rangle$.
Solid squares: Average counts of 30 simulations;
solid line: least square fit of the average counts to a sinusoidal function;
solid circles: counts obtained from individual runs with approximately $\langle N \rangle$ detected neutrons;
error bars: one standard deviation.
All simulations where carried out with $R=0.2$ and $\gamma=0.5$.
}%
\label{figure.10}
\end{center}
\end{figure*}

\section{Time-dependent phenomena}\label{time}

An important question is whether the event-based approach
leads to new predictions that may be tested experimentally and may therefore reveal new physics.
As demonstrated in this paper, and in our earlier work on (quantum) optics experiments
(for a review see Ref.~\cite{MICH11a},
in the stationary state (meaning after processing many events)
the event-based simulation reproduces the statistical distributions of quantum theory.

Therefore, new predictions can only appear when the event-based simulation
is operating in the transient regime, before the processors reach their stationary state.
Optics experiments with a Mach-Zehnder interferometer and two-beam interference that may be able
to address this question have been discussed in Ref.~\cite{MICH12a} and Ref.~\cite{JIN10b}, respectively.
Neutron interferometry is well-suited to address this issue because
the neutrons can be detected with almost 100\% efficiency and
a relatively low flux of neutrons facilitates labeling each detection event by a time stamp.
Therefore, it is feasible to collect detailed information about each neutron, which can then be analyzed further.

\subsection{Low-counting-rate experiments}\label{counting}

In these experiments, interference patterns were recorded such that in a fixed time frame,
the sum of neutron counts in the O-beam over all chosen settings of the phase shifts is approximately
constant (and relatively small)~\cite{RAUC90,ZAWI93,RAUC00}.
This is accomplished by increasing the rotation angle $\chi$ of the phase shifter at regular
time intervals.

Adopting the same procedure as in experiment, the event-based simulation
yields the results presented in Fig.~\ref{figure.10}.
The simulation data show the same qualitative features as found in experiment~\cite{RAUC90,RAUC00}.

It is instructive to inquire why the event-based processor is able to reproduce all these features,
even though it operates far from a stationary-state regime.
The main reason for this can be traced back to the speed with which the DLM responds to a change in the input messages
(see section~\ref{DLM}).
In Section~\ref{BS}, we explained that $\gamma$ controls the speed and accuracy with which
the event-based processor responds to a change in the ratio of the input events on its input ports.
In the case of the neutron interferometer (see Fig.~\ref{MZIexperiment}),
the ratio of the number of neutrons that travels from BS1 to BS3
and from BS2 to BS3 is independent of the phase shift $\chi$.
Therefore, as far as the response time to a change is concerned, only the message content matters.
However, the event-based processor described in Section~\ref{BS}
is constructed such that the last messages that were delivered at
input ports 0 and 1 are stored in the DLM.
Hence, the response to a change in one of the messages is immediate and without introducing errors.

\begin{figure*}[t]
\begin{center}
\includegraphics[width=14cm ]{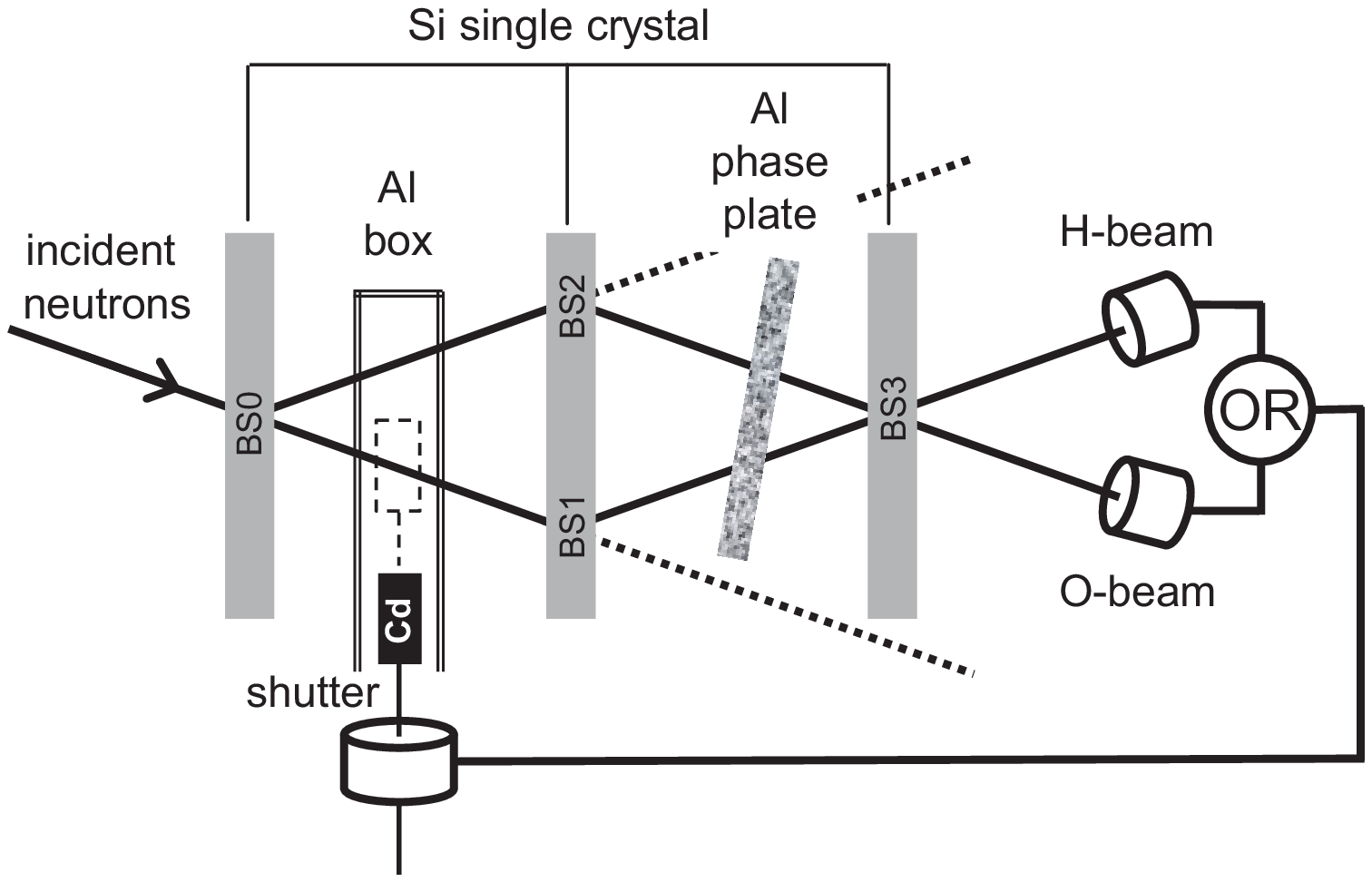}
\caption{%
Layout of Summhammer's neutron interferometry experiment with a shutter~\cite{SUMM89}.
BS0,...,BS3: beam splitters.
Neutrons which are transmitted by BS0 may be blocked by a piece of Cd metal, depending
on the state of the shutter.
Phase shifter: aluminum foil.
Neutrons that are transmitted by BS1 or BS2 leave the interferometer
and do not contribute to the neutron counts in the O- or H-beam.
For each detected neutron, the state of the shutter changes with probability 1/2~\cite{SUMM89}.
The detection events are labeled by the state of the shutter.
}%
\label{SUMMexperiment}
\end{center}
\end{figure*}

\subsection{Time-dependent beam blocking}\label{shutter}
\begin{figure*}[t]
\begin{center}
\includegraphics[width=5.5cm ]{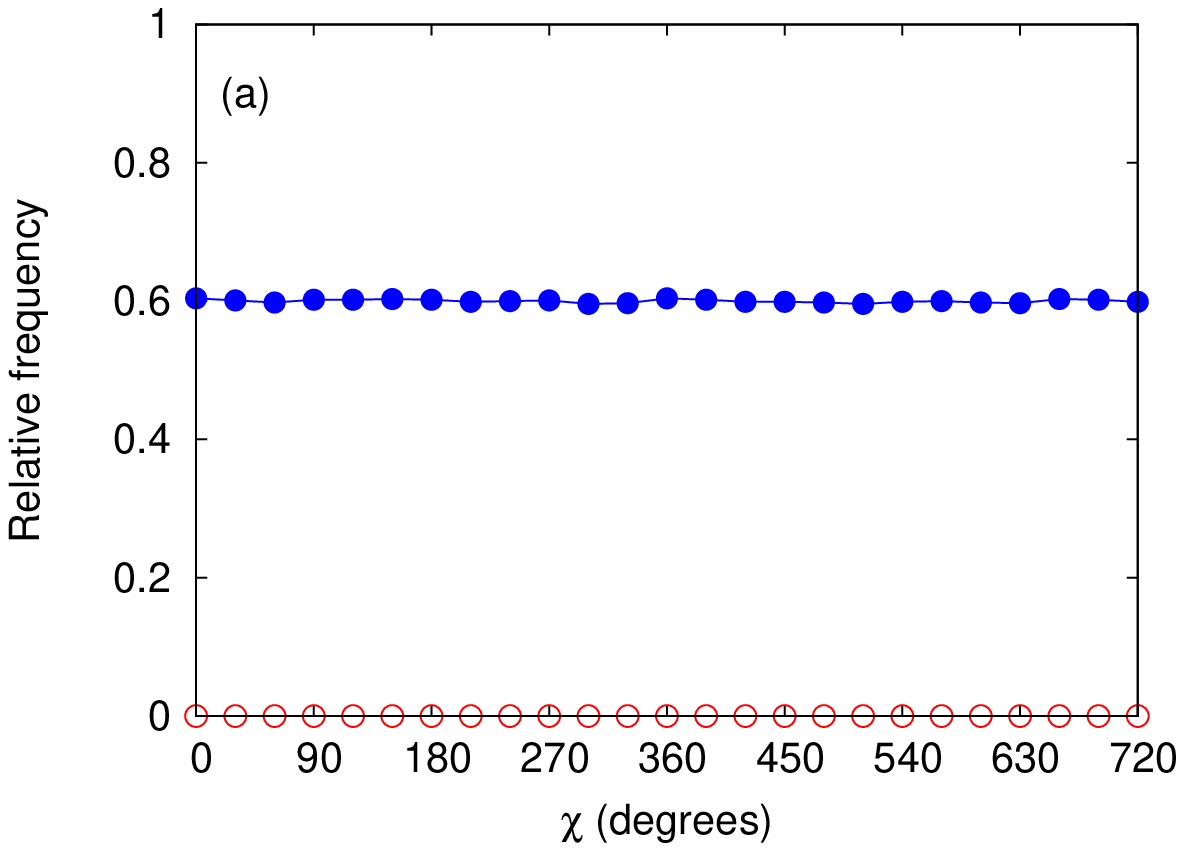}
\includegraphics[width=5.5cm ]{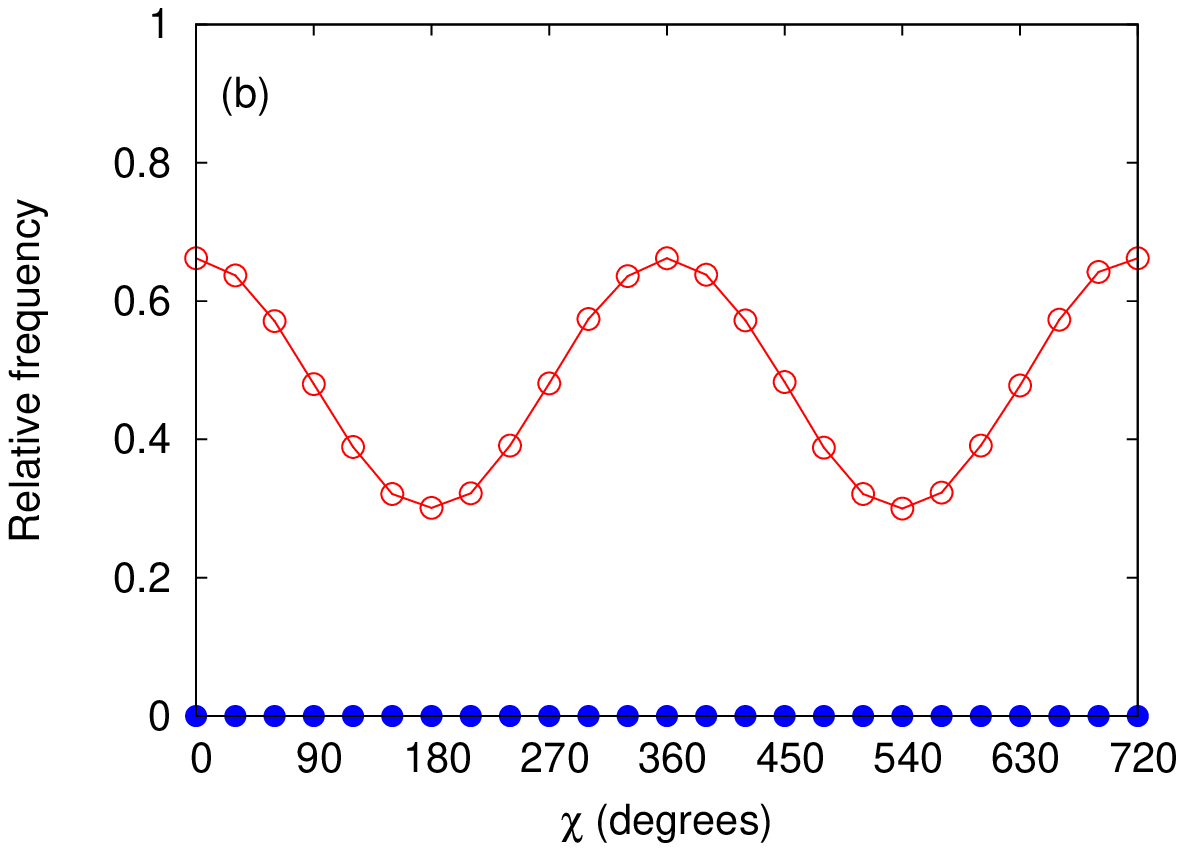}
\includegraphics[width=5.5cm ]{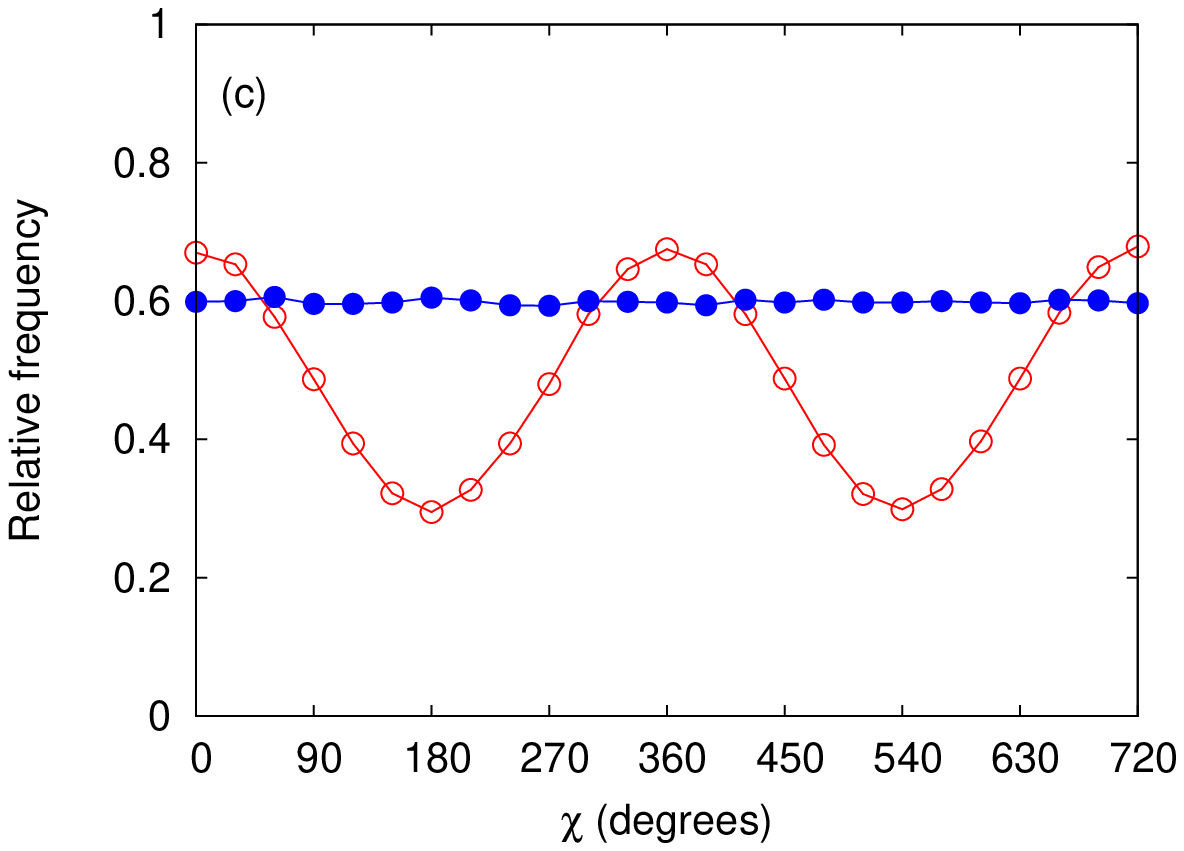}
\caption{%
Event-based simulation results of Summhammer's neutron interferometry experiment with a time-dependent shutter~\cite{SUMM89},
see Fig.~\ref{SUMMexperiment}.
The relative frequency is the neutron count in the O-beam divided by the sum of the counts in the O- and H-beam.
Open (solid) circles: relative frequency of events recorded with the shutter open (closed).
a) Shutter closed;
b) Shutter open;
c) For each neutron detected in the O- or H-beam, the state of the shutter changes with probability 1/2.
Simulation parameters:
number of incident neutrons $N=250000$, reflection $R=0.4$, and $\gamma=0.12$.
In case (c), the total number of events (per value of $\chi$) registered in the O- and H-beam is approximately 60000,
as in experiment~\cite{SUMM89}.
}%
\label{figure.11}
\end{center}
\end{figure*}
\begin{figure*}[t]
\begin{center}
\includegraphics[width=5.5cm ]{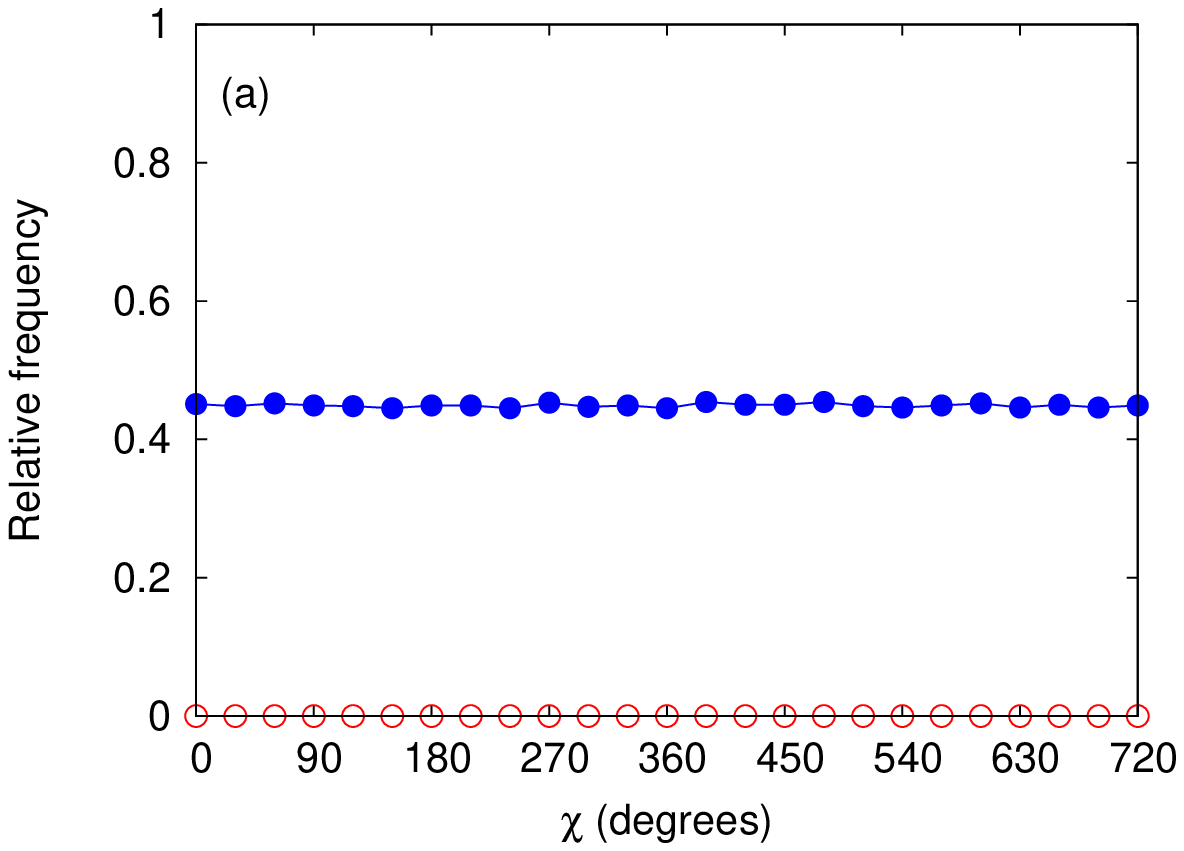}
\includegraphics[width=5.5cm ]{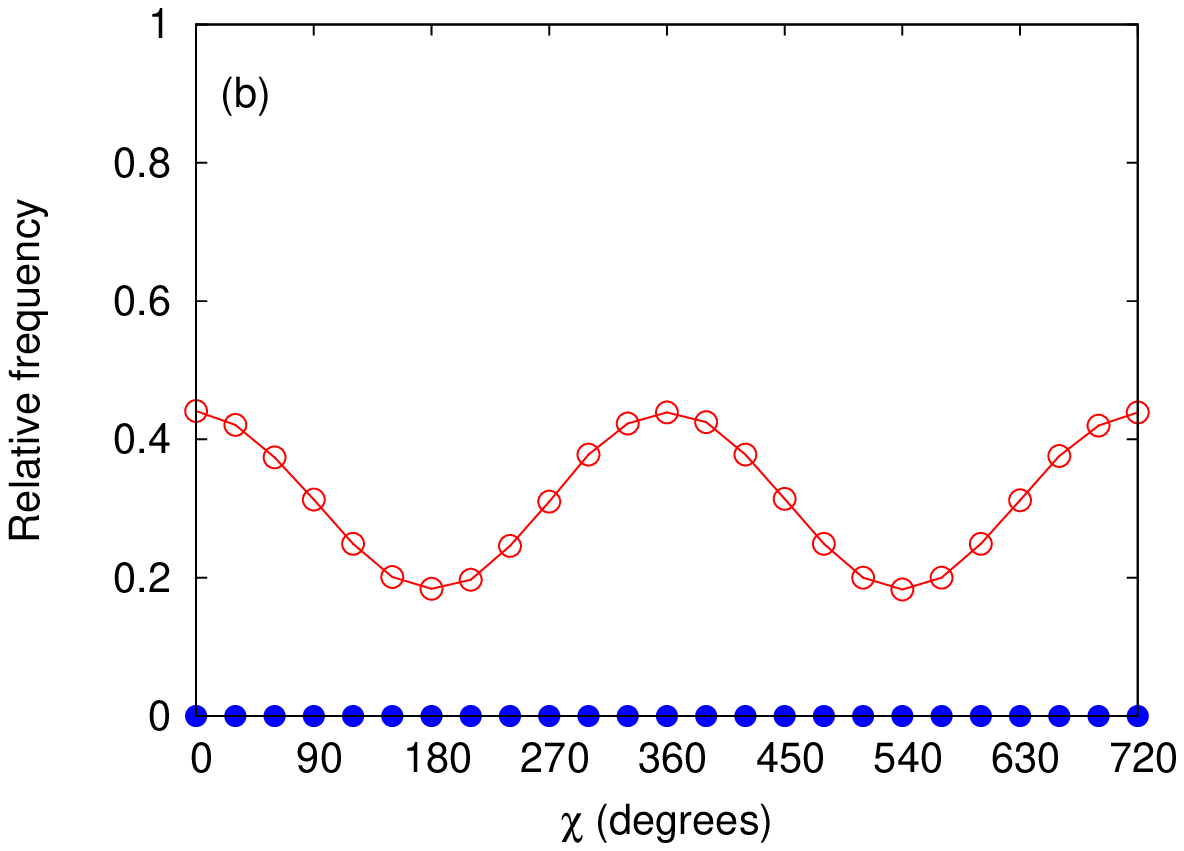}
\includegraphics[width=5.5cm ]{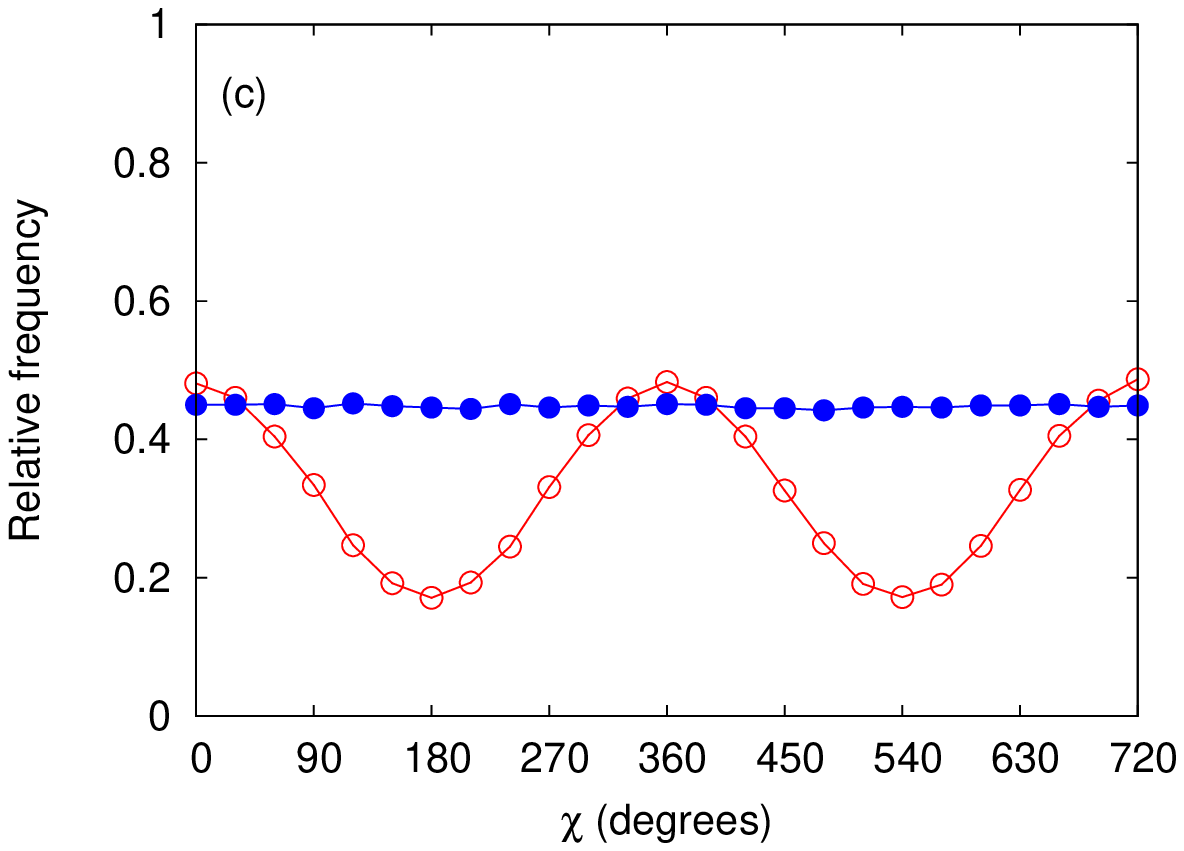}
\caption{%
Same as Fig.~\ref{figure.11} using the simplest approximation to account
for the incoherence of the incident neutron beam (see text).
Simulation parameters:
number of incident neutrons $N=500000$, reflections $R_1=0.2$, $R_2=0.9$, weights $W_1=20/21$, $W_2=1/21$, and $\gamma=0.12$.
In case (c), the total number of events (per value of $\chi$) registered in the O- and H-beam is approximately 60000,
as in experiment~\cite{SUMM89}.
}%
\label{figure.12}
\end{center}
\end{figure*}

Summhammer's neutron interferometry experiment~\cite{SUMM89} with a shutter which, upon detection of a neutron, randomly blocks one of the paths
through the interferometer provides what is perhaps the most stringent test of the event-based
model which we present in this paper.

The experimental setup is sketched in Fig.~\ref{SUMMexperiment}.
Depending on the state of the Cd metal shutter, neutrons transmitted by BS0 are blocked.
Neutrons transmitted by BS1 or BS2 leave the interferometer
and do not contribute to the neutron counts in the O- or H-beam.
Upon detection of a neutron, the shutter may change its state with probability 1/2.
Detection events are labeled by the state of the shutter.

The experimental data show the following features (see Fig.3 of Ref.~\cite{SUMM89}):
\begin{enumerate}
\item{If the shutter is kept closed, the relative frequency of the O-beam detection events
does not depend on the phase shift $\chi$, that is there is no interference pattern.}
\item{If the shutter is kept open, the relative frequency of the O-beam detection events
shows the sinusoidal dependence on $\chi$, the characteristic signature of interference.}
\item{If the state of the shutter is allowed to change according to the procedure
described earlier, the relative frequency of the O-beam detection events
conditioned on the state of the shutter exhibits
a sinusoidal dependence on $\chi$ when the shutter was open and
no dependence on $\chi$ when the shutter was closed.}
\item{The maximum of the relative frequency of the O-beam detection events
when the shutter was open is approximately equal
to the relative frequency of the O-beam detection events
when the shutter was closed. This maximum is approximately 0.43.}
\item{The visibility of the relative frequency of the O-beam detection events
when the shutter was open is about 0.4.}
\end{enumerate}

Originally conceived to test a nonergodic interpretation of quantum theory~\cite{BUON80,BUON85,BUON86},
the experimental results were interpreted as being in accordance with the
Copenhagen interpretation and ruling against predictions based on the nonergodic interpretation of quantum theory~\cite{SUMM89,BUON89}.
Citing Summhammer about the latter, ``This prediction expects that in a neutron interferometer successive neutrons interact
with each other through a hypothesized memory to which each neutron contributes a little,
such that the quantum mechanical interference phenomenon only arises
when sufficiently many neutrons have passed the interferometer in
a constant experimental condition''.

As explained in Section~\ref{DLM}, the key feature of the event-based model of a beam splitter is that
it can adapt to changes of the input data, in other words, it can learn.
Obviously, learning requires some form of memory,
which in our case consists of the $\mathbf{x}$ and $\mathbf{Y}$ registers, see Fig.~\ref{figmachine0}.
As Summhammer's experiment rules against the nonergodic interpretation of quantum theory~\cite{SUMM89,BUON89}
and memory is a key feature of this interpretation, one might expect that this experiment
rules out the event-based approach as well.
However, as we now show, this is not the case.

For all experiments considered in this paper, it is essential to include into the event-based
model all the essential aspects of the real experiments,
not just those that are considered relevant on the level of idealized thought-experiments.
In Summhammer's experiment, the motion of the shutter induces vibrations in the silicon crystal~\cite{SUMM89,BUON89}.
In the absence of concrete knowledge about this effect,
it seems very difficult to model in detail how the opening and closing of the shutter
affects the vibrating crystal.
Therefore, we adopt a pragmatic approach to mimic the effect of the shutter motion on the interferometer.
Of course, we could try different update rules for the event-based processors
but even that is not necessary.
From computer experiments, we found that it is sufficient to reset the
internal vectors $\mathbf{x}$ to zero each time the shutter closes.
Alternatively, we can set
their respective values of $\gamma$ to zero each time the shutter closes
and reset $\gamma$ of a particular processor to its specified value after
it has sent out a message.
Both these modifications have a simple physical interpretation
in terms of shaking the crystal and temporarily destroying the coherence
in (parts of) the silicon crystal.
Obviously, the ability to easily incorporate the effect of such processes
is a powerful feature of a discrete-event simulation approach.

A first set of simulation results is shown in Fig.~\ref{figure.11}.
It is clear that the event-based simulation reproduces the main
features (see items 1-3 above) of Summhammer's experimental results~\cite{SUMM89}.
What is still missing is the quantitative agreement with Summhammer's data.
In fact, we have found it to be impossible to use the parameters $R$ and $\gamma$
to achieve good numerical agreement.
This simply means that we should consider moving away from
the effective, average reflection coefficient characterization
of the beam splitter, in agreement with the theory
given in Ref.~\cite{RAUC00}.

Instead of using one reflection coefficient $R$,
let us try to use only two reflection coefficients $R_1$ and $R_2$ (for BS0,...,BS3).
Two corresponding weights $W_1$ and $W_2$ determine the frequency
with which the individual neutron ``sees'' one of the two reflection coefficients.
This approximation can be viewed as a minimalistic substitute of the incoherent averaging
over the Pendell\"osing oscillations and misset angle
in the wave theoretical treatment of the perfect crystal neutron
interferometer (see Chapter 10 of Ref.~\cite{RAUC00}).
As Fig.~\ref{figure.12} demonstrates, this minimal extension
suffices to reach quantitative agreement with Summhammer's data,
that is the simulation reproduces all five features listed above.

\subsubsection*{Quantum theoretical treatment}

It is instructive to scrutinze the statement that the experimental results
are in accordance with the Copenhagen interpretation of quantum theory~\cite{SUMM89}.
Adopting the effective, one-parameter model of the beam splitter,
in the case that the shutter is open, quantum theory predicts that
the probabilities to observe a neutron are given by (see Eq.~(\ref{app2}))
\begin{eqnarray}
p_\mathrm{O}^\mathrm{open}&=&2TR^2\left( 1+ v\cos\chi\right)
\nonumber \\
p_\mathrm{H}^\mathrm{open}&=&R\left(T^2+R^2-2vRT\cos\chi\right)
\nonumber \\
p_\mathrm{BS1}^\mathrm{open}&=&T^2
\nonumber \\
p_\mathrm{BS2}^\mathrm{open}&=&TR
,
\label{summ0}
\end{eqnarray}
where  $p_\mathrm{BS1}^\mathrm{open}$ and $p_\mathrm{BS2}^\mathrm{open}$ are the probabilities that
the neutrons are transmitted by BS1 and BS2, respectively.
Anticipating for the observation that the visibility of the interference fringes
is (much) less than one, we have introduced the visibility $v$ as an adjustable parameter.
On the other hand, if the shutter is closed we have
\begin{eqnarray}
p_\mathrm{O}^\mathrm{closed}&=&TR^2
\nonumber \\
p_\mathrm{H}^\mathrm{closed}&=&R^3
\nonumber \\
p_\mathrm{Cd}^\mathrm{closed}&=&T
\nonumber \\
p_\mathrm{BS2}^\mathrm{closed}&=&RT
,
\label{summ1}
\end{eqnarray}
where  $p_\mathrm{Cd}^\mathrm{closed}$ represents the probability
that the neutron is absorbed by the Cd shutter.

From Eqs.~(\ref{summ0}) and (\ref{summ1}), it follows that the relative frequencies to observe
neutrons in the O-beam are given by
\begin{eqnarray}
f^\mathrm{open}(\chi)&=&\frac{p_\mathrm{O}^\mathrm{open}}{p_\mathrm{O}^\mathrm{open}+p_\mathrm{H}^\mathrm{open}}=2TR\left(1+v\cos\chi\right),
\\
f^\mathrm{closed}&=&\frac{p_\mathrm{O}^\mathrm{closed}}{p_\mathrm{O}^\mathrm{closed}+p_\mathrm{H}^\mathrm{closed}}=T
.
\label{summ2}
\end{eqnarray}
The experiment shows that $\max_\chi f^\mathrm{open}(\chi)=f^\mathrm{closed}$ (see (4) in the list above)
hence $2(1+v)TR=T$ or, using the observation that $v\approx0.4$ (see (5) in the list above),
$R=1/2(1+v)\approx0.36$ such that $f^\mathrm{closed}\approx0.64$, which is
incompatable with the experimental observation that $f^\mathrm{closed}\approx0.43$ (see (4) in the list above).
Therefore, the experimental data rules out the two-parameter ($R,v$) quantum model of this interferometry experiment.
However, just as we did in the case of the event-based model for this experiment,
we may generalize the model summing over incoherent processes.
As before, for simplicity, we consider a model with
only two reflections $R_1$ and $R_2$ occurring with probabilities $W_1$ and $W_2=1-W_1$, respectively.

Instead of Eq.~(\ref{summ2}), we now have
\begin{eqnarray}
f^\mathrm{open}&=&2\frac{W_1T_1R_1^2+W_2T_2R_2^2}{W_1R_1+W_2R_2}\left( 1+ v\cos\chi\right)
\nonumber \\
f^\mathrm{closed}&=&\frac{W_1T_1R_1^2+W_2T_2R_2^2}{W_1R_1^2+W_2R_2^2}
,
\label{summ3}
\end{eqnarray}
and imposing the conditions
$\max_\chi f^\mathrm{open}(\chi)=f^\mathrm{closed}=g$ yields
\begin{eqnarray}
2(1+v)(W_1R_1^2+W_2R_2^2)&=&W_1R_1+W_2R_2,
\nonumber \\
W_1T_1R_1^2+W_2T_2R_2^2&=&g(W_1R_1^2+W_2R_2^2)
.
\label{summ4}
\end{eqnarray}
These equations can be solved in closed form, the non-trivial solution reading
\begin{eqnarray}
R_2&=&\frac{1-g-R_1}{1-2(1+v)R_1},
\nonumber \\
W_1&=&
\frac{(1-g-R_1) (-2 v+2 g (v+1)-1)}{8 (v+1)^3 R_1^4-12 (v+1)^2 R_1^3+6 (v+1) R_1^2-2 (g+(g-1) v) R_1-(g-1) (-2 v+2 g (v+1)-1)}
.
\label{summ5}
\end{eqnarray}
Note that the solutions are bounded by $0<R_1<1$, $0<R_2<1$, $0\le v\le 1$, $0\le g \le1$ and $0<W_1<1$.

Taking for instance $R_1=0.2$, $v=0.4$ and $g=0.43$, we obtain $R_2=0.84$ and $W_1=1-W_2=0.93$, which are all
reasonable numbers, the values of $R_1,R_2,W_1$ and $W_2$ being rather close to the numbers that
were used in the event-based simulation (see Fig.~\ref{figure.12}).
In conclusion, it is clear that the four-parameter ($R_1,R_2,W_1,v$) quantum model is compatible with
the experimental data.
Note that this compatibility is not due to some unique feature of quantum theory
but merely results from adding, with appropriate weights,
the results of two independent experiments performed under different conditions.

In spite of the fact that quantum theory can be used to describe the outcome
of Summhammer's experiment, the ``mystery'' alluded to in the introduction remains.
First, it is mysterious how the experimental apparatus can ``know''
the expresssions of the probabilities Eqs.~(\ref{summ0}) and (\ref{summ1})
before the very first neutron has passed through it and a decision
about the state of the shutter was taken.
Second, quantum theory postulates that a detection event corresponds to
a certain value (one out of four in this case) of a random variable
with a probability distribution given by either
Eq.~(\ref{summ0}) or Eq.~(\ref{summ1})
but (like probability theory in general) is silent
about the process by which these random values are actually realized.
There are no such mysteries in the event-based approach as it provides a complete prescription
of how individual (detection) events are to be generated.

\section{Summary and outlook}\label{summary}

In this paper, we have demonstrated that the event-based approach, originally
introduced in Ref.~\cite{RAED05b,RAED05c,RAED05d} to simulate
quantum optics experiments, can also be applied to neutron interferometry experiments.
Our approach gives a detailed, mystery-free, particle-only description of interference and entanglement,
as observed in neutron interferometry experiments~\cite{RAUC00} and
does not suffer from the quantum measurement problem, simply because
the discrete events, such as the detection of a neutron, are taken as the basic
entities of the description.
The statistical distributions which are observed in real experiments, usually thought
to be of quantum mechanical origin, emerge from a time series of discrete events
generated by causal, classical, adaptive systems.

Conceptually, our approach may seem to have similarities to cellular autonoma modeling~\cite{FRED90,WOLF02},
the work of 't Hooft~\cite{THOO07,THOO12} or to, for instance,
lattice Boltzmann modeling of fluid dynamics~\cite{SUCC01}
which all explore the idea that simple rules,
which not necessarily derive from classical Hamiltonian dynamics,
define discrete-event processes which may lead to
the (complicated) behavior that is observed in experiments.
However, the reasoning that lead us to our simulation model is very different.

Starting from the point of view that empirical knowledge,
and the concepts created on the basis of this knowledge,
derives from the elementary events which are registered by our senses,
we explore the consequences of assuming that current scientific knowledge is built
on the notion of discrete events and the relations between them.
This is a departure from the prevailing mode of thinking
in theoretical physics, which assumes that the definite results which we observe
are signatures of an underlying objective reality
that is mathematical in nature.
While the hypothesis that such a reality exists cannot be refuted on logical grounds,
their is no experimental evidence that supports it.
Taking an indifferent stance on this issue, the urge to ``deduce'' the existence of
definite results (discrete events) from a set of axioms disappears,
opening a route to a mode of thinking that is much less constrained.
Apparently, this change of paradigm facilitates the construction of simulation models
which reproduce the experimental
and quantum theoretical results of many real experiments,
in particular those in which the data is recorded event-by-event.

The discrete-event model that has been described in this paper
can also be used to simulate neutron spin-echo experiments~\cite{RAUC00,GRIG04}
and the recent experiments on the uncertainty in neutron spin measurements~\cite{ERHA12}.
However, neutron (and optics) experiments that involve diffraction/scattering
cannot yet be simulated with the present model.
Incorporating this feature is left for future research.

We hope that our work will stimulate the design of new single-neutron experiments
to explore the applicability of event-based modeling to physical phenomena.
Specifically, to test the validity of our discrete-event modeling,
it would be worthwhile to repeat Summhammer's experiment with the shutter
under conditions that show much larger visibility.

\section{Acknowledgement}
We are grateful to Profs. H. Rauch and J. Summhammer for very stimulating discussions
and Dr. H. Lemmel for making experimental data available to us.
We are indebted to K. De Raedt for his help in solving the time-dependent beam blocking problem
and to Profs. M. Novotny and D. Stavenga for a critical reading of the manuscript.
This work is partially supported by NCF, the Netherlands.

\bibliography{c:/d/papers/epr11,c:/d/papers/neutrons}   

\end{document}